\documentclass{article}
\usepackage{graphicx}
\usepackage{textcomp,amsmath,amssymb,soul}
\usepackage{amsfonts}
\allowdisplaybreaks

\begin{document}

\begin{center}
{\Large\bf Pseudotensor applied to Numerical Relativity\\
in Calculating Global Quantities}
\end{center}

\begin{center}
Chung-Chin Tsai$^1$, Zhoujian Cao$^2$, Chun-Yu Lin$^3$,
and Hwei-Jang Yo$^4$\\
$^{1}$Department of Physics, National ChangHua University of Education,
ChangHua City, Taiwan, ROC,\\
$^{2}$Institute of Applied Mathematics and LSEC, Academy of
Mathematics and Systems Science, Chinese Academy of Sciences,
Beijing 100190, China,\\
$^{3}$National Center for High-Performance Computing, Hsinchu 300, Taiwan,
ROC,\\
$^4$Department of Physics, National Cheng-Kung University, Tainan City,
Taiwan, ROC.
\end{center}

\begin{abstract}
In this work we apply the Landau-Lifshitz pseudotensor flux formalism 
to the calculation of the total mass and the total angular momentum during
the evolution of a binary black hole system.
We also compare its performance with the traditional integrations for the
global quantities.
It shows that the advantage of the pseudotensor flux formalism is the
smoothness of the numerical value of the global quantities,
especially of the total angular momentum.
Although the convergence behavior of the global quantities with
the pseudotensor flux method is only comparable with the ones with
the traditional method, the smoothness of its numerical value allows using
a larger radius for surface integration to obtain more accurate result.\\
$\quad$\\
{\bf This work is dedicated to the General Relativity special issue.}
\end{abstract}
\smallskip
\section{Introduction}
Since the breakthrough by Pretorius \cite{pref05} in 2005, the inspiral,
merger, and ringdown of binary compact objects (including binary black holes,
i.e., BBHs,
binary neutron stars, black hole-neutron star binary) has been successfully
simulated to high accuracy.
The emphasis in the field is now turned to extracting 
astrophysical information from these simulations.
Therefore, it will be useful in validation to measure one physical quantity
with different methods.

The pseudotensor formalisms \cite{lali62} and the quasilocal quantities
\cite{szal04,ccnt05} arise from viewing general relativity as a nonlinear
field theory in a fixed background reference, especially in a flat
auxiliary spacetime.
These formalisms have been used to explore the nonlinear dynamics of spacetime,
for example, dynamical horizons \cite{aakb04}, the distribution and flow of
linear momentum in stronly nonlinearly curved spacetimes \cite{kdea09}, and
black hole spin measurement \cite{mvfm15}.
The result shows the usefulness of these formalisms and also shed a light
on their possibly broader applications as analytical tools in various
numerical simulations.

In this work we would like to try an alternative method in calculating the
total mass and the total angular momentum with the Landau-Lifshitz
pseudotensor formalism.
The motivation comes from the imperfection of the commonly used method,
i.e., Eqs.~(\ref{surfmass}) and (\ref{surfang}),
in calculating the Arnowitt-Deser-Misner (ADM) mass and the angular momentum.
The calculation of these two global quantities basically executes an
integration over a two-sphere on the spatial domain.
Due to the limit of the grid resolution, the numerical values of these
two quantities, especially the one of the angular momentum, fluctuate
around the average values.
The numerical fluctuation makes it difficult to tell the numerical value
accurately.
Different from the ADM mass and the angular momentum,
the calculation of the momentum flux from the Landau-Lifshitz pseudotensor
formalism includes not only an integration over a two-sphere on the spatial
domain, but also an integration over the time domain.
We expect that this method with the extra integration will give smoother
global quantity curves with respect to time, and at least as accurate as
the one for the ADM mass and the angular momentum.

The rest of this work is organized as follows:
In the next section, we give a description of the
Baumgarte-Shapiro-Shibata-Nakamura (BSSN) formulation commonly used in
numerical relativity.
We then describe the methods used in this work for the calculation of the
global quantities, i.e., the total mass and the angular momentum,
in Sec.~\ref{secIII}.
In this section, besides the usual ADM and the angular momentum calculation,
it gives the integral formulas of these global quantities with the
Landau-Lifshtiz pseudotensor.
We then report the numerical comparison between these two different methods
in the simulation of BBH with spins in Sec.~\ref{secIV}.
And the discussion and summary will be presented in the Sec.~\ref{secV}.
Throughout the paper, geometric units with $G=c=1$ are used.
Einstein summation rule is adopted unless stated explicitly.

\section{The BSSN Formulation}\label{bssnf}
The metric in the ADM form is
\begin{equation} \label{adm_metric}
{\rm d}s^2=-\alpha^2{\rm d}t^2+\gamma_{ij}({\rm d}x^i+\beta^i{\rm d}t)
({\rm d}x^j+\beta^j{\rm d}t),
\end{equation}
wherein $\alpha$ is the lapse function, $\beta^i$ is the shift vector,
and $\gamma_{ij}$ is the spatial three-metric.  Throughout this paper, Latin
indices are spatial indices and run from 1 to 3, whereas Greek indices
are space-time indices and run from 0 to 3.

Einstein's equations can then be decomposed into the Hamiltonian
constraint ${\mathcal H}$ and the momentum constraints ${\mathcal
M}_i$
\begin{align}
   {\mathcal H} &\equiv R - K_{ij}K^{ij} + K^2 = 0, \label{ham1} \\
   {\mathcal M}_i&\equiv \nabla_j K^{j}_{~i} - \nabla_i K = 0, \label{mom1}
\end{align}
and the evolution equations
\begin{align}
\frac{\rm d}{{\rm d}t} \gamma_{ij} & =- 2 \alpha K_{ij},\label{gdot1} \\
\frac{\rm d}{{\rm d}t}K_{ij}&=-\nabla_i \nabla_j\alpha+\alpha( R_{ij}
        - 2 K_{i\ell} K^\ell_{~j} + K K_{ij}).       \label{Kdot1}
\end{align}
Here we have assumed vacuum $T_{\alpha\beta} = 0$ and have used
\begin{equation}
\frac{\rm d}{{\rm d}t} = \frac{\partial}{\partial t} - \pounds_{\vec\beta},
\end{equation}
where $\pounds_{\vec\beta}$ is the Lie derivative with respect to $\beta^i$.
$\nabla_i$ is the covariant derivative associated with
$\gamma_{ij}$, $R_{ij}$ is the three-dimensional Ricci tensor
\begin{equation} \label{ricci}
   R_{ij}=\frac{1}{2} \gamma^{k\ell} \left( \gamma_{kj,i\ell}
    +\gamma_{i\ell,kj}-\gamma_{k\ell,ij}-\gamma_{ij,k\ell}\right)
    + \Gamma^{mk}{}_i\Gamma_{mkj} - \Gamma_{mij} \Gamma^{mk}{}_k,
\end{equation}
where
\begin{equation}
\Gamma^{i}{}_{jk}\equiv\frac{1}{2}\gamma^{i\ell}(\gamma_{\ell j,k}+
\gamma_{\ell k,j}-\gamma_{jk,\ell}).
\end{equation}
And $R$ is its trace $R = \gamma^{ij} R_{ij}$.

In the BSSN formalism \cite{BSSN95_99},
the above ADM equations are rewritten by introducing the conformally related
metric $\tilde \gamma_{ij}$
\begin{equation}\label{confgij}
\tilde \gamma_{ij} = e^{- 4 \phi} \gamma_{ij},
\end{equation}
with the conformal exponent $\phi$ chosen so that the determinant
$\tilde\gamma$ of $\tilde \gamma_{ij}$ is unity
\begin{equation}
e^{4 \phi} = \gamma^{1/3},
\end{equation}
where $\gamma$ is the determinant of $\gamma_{ij}$.
The traceless part of the extrinsic curvature $K_{ij}$, defined by
\begin{equation}\label{tracelessK}
A_{ij} = K_{\langle ij\rangle}\equiv K_{ij} - \frac{1}{3} \gamma_{ij} K,
\end{equation}
where $K_{ij}$ with two indices between $\langle\rangle$ is to take the
symmetric and traceless part of $K_{ij}$, and
$K = \gamma^{ij} K_{ij}$ is the trace of the extrinsic curvature,
is conformally decomposed according to
\begin{equation}
\tilde A_{ij} = e^{- 4 \phi} A_{ij}.
\end{equation}
The conformal connection functions $\tilde{\Gamma}^i$, initially defined as
\begin{equation} \label{Gamma}
\tilde{\Gamma}^i \equiv \tilde{\gamma}^{jk} \tilde{\Gamma}^i_{jk}
        = - \tilde{\gamma}^{ij}_{~~,j},
\end{equation}
are regarded as independent variables in this formulation.

The evolution equations of BSSN formulation can be written as
\begin{align}
\frac{\rm d}{{\rm d}t}\phi&=-\frac{1}{6}\alpha K,\label{eq:evolphi}\\
\frac{\rm d}{{\rm d}t}\tilde{\gamma}_{ij}&=-2\alpha\tilde{A}_{ij},\label{dtg}\\
\frac{\rm d}{{\rm d}t}K&=\alpha\left(\tilde{A}_{ij}\tilde{A}^{ij}
+\frac{1}{3}K^2\right)-\nabla^2\alpha, \label{dtK}\\
\frac{\rm d}{{\rm d}t}\tilde{A}_{ij}&=
        \alpha(K\tilde{A}_{ij}-2\tilde{A}_{ik}\tilde{A}^k{}_j)+
e^{-4\phi}(\alpha R_{\langle ij\rangle}-\nabla_{\langle i}\nabla_{j\rangle}
\alpha),\label{dtA}\\
\partial_t\tilde{\Gamma}^i&=2\alpha\left(\tilde{\Gamma}^i_{jk}
    \tilde A^{jk}-\frac{2}{3}\tilde{\gamma}^{ij}K_{,j}
    +6\tilde{A}^{ij}\phi_{,j}\right)-2\tilde{A}^{ij}\alpha_{,j}\nonumber\\
   &+\beta^j\tilde{\Gamma}^i{}_{,j}-\tilde{\Gamma}^j\beta^i{}_{,j}
+\frac{2}{3}\tilde{\Gamma}^i\beta^j{}_{,j}
+\tilde{\gamma}^{jk}\beta^i{}_{,jk}+
  \frac{1}{3}\tilde{\gamma}^{ij}\beta^k{}_{,jk}.\label{dtGamma}
\end{align}
The Ricci tensor $R_{ij}$ can be written as a sum of two pieces
\begin{equation}
   R_{ij} = \tilde{R}_{ij} + R^{\phi}_{ij},
\end{equation}
where $R^{\phi}_{ij}$ is given by
\begin{equation}
   R^{\phi}_{ij}=-2\tilde{\nabla}_i\tilde{\nabla}_j\phi-2\tilde{\gamma}_{ij}
    \tilde{\nabla}^2\phi+4\tilde{\nabla}_i\phi\tilde{\nabla}_j\phi
   -4\tilde{\gamma}_{ij}\tilde{\nabla}^k\phi\tilde{\nabla}_k\phi,
\end{equation}
where ${\tilde\nabla}_i$ is the covariant derivative with respect to
${\tilde\gamma}_{ij}$,
while, with the help of the $\tilde \Gamma^i$, $\tilde{R}_{ij}$
can be expressed as
\begin{equation}
   \tilde{R}_{ij}=-\frac{1}{2}\tilde{\gamma}^{mn}\tilde{\gamma}_{ij,mn}
     +\tilde{\gamma}_{k(i}\tilde{\Gamma}^k{}_{,j)}
     +\tilde{\Gamma}^k\tilde{\Gamma}_{(ij)k}
    +2\tilde{\Gamma}^{k\ell}{}_{(i}\tilde{\Gamma}_{j)k\ell}
     +\tilde{\Gamma}^{k\ell}{}_i\tilde{\Gamma}_{k\ell j}.\label{confricci}
\end{equation}
The new variables are tensor densities, so that their Lie derivatives are
\begin{align}
   \pounds_{\vec\beta}K&=\beta^kK_{,k},\\
   \pounds_{\vec\beta}\phi&=\beta^k\phi_{,k} + \frac{1}{6}\beta^k{}_{,k},\\
   \pounds_{\vec\beta}\tilde{\gamma}_{ij}&=\beta^k\tilde{\gamma}_{ij,k}
 +2\tilde{\gamma}_{k(i}\beta^k{}_{,j)}-\frac{2}{3}\tilde{\gamma}_{ij}
  \beta^k{}_{,k},\label{lieg}\\
   \pounds_{\vec\beta}\tilde{A}_{ij}&=\beta^k\tilde{A}_{ij,k}
    +2\tilde{A}_{k(i}\beta^k{}_{,j)}-\frac{2}{3}\tilde{A}_{ij}\beta^k{}_{,k}.
\end{align}
The Hamiltonian and momentum constraints (\ref{ham1}) and (\ref{mom1})
can be rewritten as
\begin{align}
&{\mathcal H}=e^{-4\phi}(\tilde{R}-8\tilde{\nabla}^2\phi
  -8\tilde{\nabla}^i\phi\tilde{\nabla}_i\phi)
   +\frac{2}{3}K^2-\tilde{A}_{ij}\tilde{A}^{ij}=0,\label{ham2}\\
&{\mathcal M}_i=\tilde{\nabla}_j\tilde{A}_i{}^j+6\phi_{,j}\tilde{A}_i{}^j-
        \frac{2}{3}K_{,i}=0,\label{mom2}
\end{align}
where $\tilde{R}=\tilde{\gamma}^{ij}\tilde{R}_{ij}$.
\section{Calculation of Global Quantities}\label{secIII}
\subsection{ADM mass and angular momentum}
The ADM mass is defined in terms of a surface integral at spatial infinity.
In numerical simulations, this integral can be approximated by an integral
evaluated on a surface near the outer boundaries of the grid.
In Cartesian coordinates, the ADM mass is defined by a surface integral at
spatial infinity \cite{MTW,mnyj74}
\begin{equation}\label{tote}
M=\frac{1}{16\pi}\oint_\infty \gamma^{im}\gamma^{jn}(\gamma_{mn,j}-
\gamma_{jn,m}){\rm d}^2\Sigma_i,
\end{equation}
where ${\rm d}^2\Sigma_i\equiv (1/2)\sqrt{\gamma}\epsilon_{ijk}{\rm d}x^j{\rm
d}x^k$ is the surface element and $\epsilon_{ijk}$ the Levi-Civita
alternating symbol.  We now perform a conformal decomposition
\begin{equation}\label{barg}
\gamma_{ij}=\psi^4\bar\gamma_{ij},
\end{equation}
where $\psi=e^\phi$.
Assuming the asymptotic behavior
\begin{equation}
\psi \sim 1 + O(\frac{1}{r}) \mbox{~~~~when~~~~} r \rightarrow \infty,
\end{equation}
and
\begin{equation}
\tilde{\gamma}_{ij} \sim \delta_{ij} + O(\frac{1}{r}) \mbox{~~~~when~~~~}
r \rightarrow \infty,
\end{equation}
we can rewrite (\ref{tote}) as
\begin{align}
   M &=\frac{1}{16\pi}\oint_\infty \psi^{-2}
    \tilde{\gamma}^{im}\tilde{\gamma}^{jn}\left[\psi^4(\tilde{\gamma}_{mn,j}
    -\tilde{\gamma}_{jn,m})+4\psi^3(\psi_{,j}\tilde{\gamma}_{mn}
    -\psi_{,m}\tilde{\gamma}_{jn})\right]{\rm d}^2\tilde{\Sigma}_i\nonumber \\
   &=\frac{1}{16\pi}\oint_\infty
    \tilde{\gamma}^{im}\left[\tilde{\gamma}^{jn}(\tilde{\gamma}_{mn,j}-
\tilde{\gamma}_{jn,m})-8\psi_{,m}\right]{\rm d}^2\tilde{\Sigma}_i\nonumber\\
   &=\frac{1}{16\pi}\oint_\infty (\tilde{\Gamma}^i-\tilde{\Gamma}^{ji}
    {}_j-8\tilde{\nabla}^i\psi){\rm d}^2\tilde{\Sigma}_i=\frac{1}{16\pi}
\oint_\infty (\tilde{\Gamma}^i-8\tilde{\nabla}^i e^\phi)
{\rm d}^2\tilde{\Sigma}_i.\label{newder}
\end{align}
Here the conformal surface element is defined as
${\rm d}^2\tilde{\Sigma}_i=(1/2)\epsilon _{ijk}{\rm d}x^j{\rm d}x^k$
since $\tilde{\gamma}=1$, we use the abbreviations
$\tilde{\Gamma}^i\equiv\tilde{\gamma}^{jk}\tilde{\Gamma}^i{}_{jk}$ and
$\tilde\Gamma^{ji}{}_j\equiv \tilde{\gamma}^{ik}\tilde{\Gamma}^j{}_{kj}=0$,
and $\tilde\nabla_i$ is the three-covariant derivative with respect to the
metric $\tilde{\gamma}_{ij}$.

We define the angular momentum $J^i$ as (compare \cite{york79,by80})
\begin{equation}\label{Jdef1}
   J_i\equiv\frac{1}{8\pi} \epsilon_{ij}{}^k \oint_\infty x^j A^\ell{}_k
{\rm d}^2\Sigma_\ell=
\frac{1}{8\pi}\epsilon_{ij}{}^k\oint_\infty x^j e^{6\phi}\tilde{A}^\ell{}_k
{\rm d}^2\tilde{\Sigma}_\ell,
\end{equation}
where the indices of $\epsilon_{ij}{}^k$ are raised and lowered with the
flat metric $\delta_{ij}$,
${\rm d}^2\Sigma_i=e^{6\phi}{\rm d}^2\tilde{\Sigma}_i$, and
$A^i{}_j=\tilde{A}^i{}_j$.

Therefore, the surface integrals of the ADM mass and the angular momentum
(in vacuum) are respectively \cite{yhbs02}:
\begin{align}
M=&\frac{1}{16\pi}\oint_{\partial\Omega}({\tilde\Gamma}^i-8\tilde{\nabla}^i
e^\phi){\rm d}^2{\tilde\Sigma}_i,\label{surfmass}\\
J_i=&\frac{1}{8\pi} \epsilon_{ij}{}^k \oint_{\partial\Omega}e^{6\phi}x^j
\tilde{A}^\ell{}_k{\rm d}^2\tilde{\Sigma}_\ell.\label{surfang}
\end{align}
These two global quantities are useful tools for the system diagnostics to
validate the calculations.
\subsection{Pseudotensor and momentum flux}
In this section, we briefly review the Landau-Lifshitz formulation of gravity
and the statement of four-momentum conservation within this theory.
The Landau-Lifshitz formulation has been described in
\cite{lali62,MTW} to reformulate general relativity as a nonlinear
field theory in flat spacetime.
Here we follow closely to the content in \cite{kdea09}.
In this formalism, an arbitrary
asymptotically Lorentz coordinate is firstly built on a given curved
 (but asymptotically-flat) spacetime.
Then the coordinate is used to map the curved (i.e. physical) spacetime onto
an auxiliary flat spacetime by enforcing that the coordinate on this spacetime
are globally Lorentz.
The auxiliary flat metric takes the Minkowski form,
$\eta_{\mu\nu}={\rm diag}(-1,1,1,1)$.

Gravity is described, in this formulation, by the physical metric density
\begin{equation}
\mathfrak{g}^{\mu \nu}\equiv\sqrt{-g} g^{\mu \nu},
\label{eqn:gg}
\end{equation}
where $g$ is the determinant of the covariant components of the physical
metric, and $g^{\mu\nu}$ are the contravariant components of the physical
metric.
In terms of the superpotential
\begin{equation}
H^{\mu\alpha\nu\beta}\equiv\mathfrak{g}^{\mu \nu}\mathfrak{g}^{\alpha \beta}
-\mathfrak{g}^{\mu \alpha}\mathfrak{g}^{\nu \beta},
\label{eqn:superpotential}
\end{equation}
the Einstein field equations take the field-theory-in-flat-spacetime form
\begin{equation}\label{eqn:efe}
H^{\mu\alpha\nu\beta}{}_{,\alpha\beta} = 16\pi \tau^{\mu\nu}.
\end{equation}
Here $\tau^{\mu\nu} = (-g)(T^{\mu\nu} + t^{\mu\nu}_{\rm LL})$ is the total
effective stress-energy tensor, indices after the comma denote partial
derivatives (covariant derivatives with respect to the flat auxiliary metric),
and the Landau-Lifshitz pseudotensor $t^{\mu\nu}_{\rm LL}$ (actually a real
tensor in the auxiliary flat spacetime) is given by
\begin{align}\label{eq:LLPseudo}
16\pi(-g)t^{\mu\nu}_{\rm LL}&=\mathfrak{g}^{\mu\nu}{}_{,\lambda}
\mathfrak{g}^{\lambda\sigma}{}_{,\sigma}-\mathfrak{g}^{\mu
\lambda}{}_{,\lambda}\mathfrak{g}^{\nu\sigma}{}_{,\sigma}+\frac{1}{2}
g^{\mu\nu}g_{\lambda\sigma}\mathfrak{g}^{\lambda\tau}{}_{,\rho}
\mathfrak{g}^{\rho\sigma}{}_{,\tau}\nonumber\\
&-2g^{(\mu|\lambda}g_{\tau\sigma}\mathfrak{g}^{|\nu)\sigma}{}_{,\rho}
\mathfrak{g}^{\tau\rho}{}_{,\lambda}
+g_{\lambda\sigma}g^{\tau\rho}\mathfrak{g}^{\mu\lambda}{}_{,\tau}
\mathfrak{g}^{\nu\sigma}{}_{,\rho}\nonumber\\
&+\frac{1}{8}(2g^{\mu\lambda}g^{\nu\sigma}-g^{\mu\nu}g^{\lambda\sigma})
(2g_{\tau\rho}g_{\kappa\eta}-g_{\rho\kappa}g_{\tau\eta})
\mathfrak{g}^{\tau\eta}{}_{,\lambda}\mathfrak{g}^{\rho\kappa}{}_{,\sigma}.
\end{align}
And the Landau-Lifshitz pseudotensor can be expressed in term of
the 4-metric $g_{\mu\nu}$ and the 4-connection
 ${\mathit\Gamma}^\mu{}_{\nu\sigma}$ as
\begin{align}
16\pi t^{\mu\nu}_{\rm LL}=&2{\mathit\Gamma}^{(\mu\nu)}{}_\sigma
({\mathit\Gamma}^\sigma-L^\sigma)+2{\mathit\Gamma}^{\sigma\mu\nu}L_\sigma
-({\mathit\Gamma}^\mu-L^\mu)({\mathit\Gamma}^\nu-L^\nu)
+{\mathit\Gamma}^\mu{}_{\lambda\sigma}{\mathit\Gamma}^{\nu\lambda\sigma}
\nonumber\\
-&2{\mathit\Gamma}^{(\mu|}{}_{\lambda\sigma}
{\mathit\Gamma}^{\lambda\sigma|\nu)}
-{\mathit\Gamma}_{\lambda\sigma}{}^\mu
{\mathit\Gamma}^{\sigma\lambda\nu}+g^{\mu\nu}(L_\sigma L^\sigma
-2L_\sigma{\mathit\Gamma}^\sigma+{\mathit\Gamma}_{\lambda\sigma\rho}
{\mathit\Gamma}^{\sigma\lambda\rho}),
\end{align}
where ${\mathit\Gamma}^\mu\equiv g^{\lambda\sigma}
{\mathit\Gamma}^\mu{}_{\lambda\sigma}$,
$L_\mu\equiv{\mathit\Gamma}^\sigma{}_{\mu\sigma}$.
With the relation equations in Appendix \ref{app1}, the equation can be
re-expressed easily with the $3+1$ quantities.
By virtue of the symmetries of the superpotential (which are the same as those
of the Riemann tensor), the field equations in the form (\ref{eqn:efe}) imply
the differential conservation law for four-momentum
\begin{equation}
{\tau^{\mu\nu}}_{,\nu} = 0,
\label{eqn:divtau}
\end{equation}
which is equivalent to ${T^{\mu\nu}}_{;\nu} = 0$ (where the semicolon denotes
a covariant derivative with respect to the physical metric).

It is shown in \cite{lali62,MTW} that the total four-momentum of any
isolated system as measured gravitationally in the asymptotically flat region
far from the system is
\begin{equation}
p^\mu_{\rm tot}=\int_V\tau^{\mu 0}{\rm d}^3 x.\label{eqn:MomTot}
\end{equation}
Thus
\begin{equation}
\frac{{\rm d}p^\mu_{\rm tot}}{{\rm d}t}=\frac{\rm d}{{\rm d}t}
\int_V\tau^{\mu 0}{\rm d}^3 x=\int_V\tau^{\mu 0}{}_{,0}{\rm d}^3 x
=-\int_V\tau^{\mu j}{}_{,j}{\rm d}^3 x
=-\oint_S\tau^{\mu j}{\rm d}^2\bar\Sigma_j\;,\label{eqn:dMomTotdt}
\end{equation}
where ${\rm d}^2\bar\Sigma_i\equiv(1/2)\epsilon_{ijk}{\rm d}x^j{\rm d}x^k$
is the surface-area element
(defined by using the flat auxiliary metric), and the integral is over an
arbitrarily large closed surface $S$ surrounding the system,
and Eq.~(\ref{eqn:divtau}) is used.
Therefore, the total momentum flux across the 2-surface within $[t_1,t_2]$ is
\begin{equation}
\Delta p^\mu_{\rm tot}\equiv p^\mu_{\rm tot}(t_2)-p^\mu_{\rm tot}(t_1)=
-\int^{t_2}_{t_1}\oint_S\tau^{\mu j}{\rm d}^2\bar\Sigma_j{\rm d}t.
\end{equation}
With $p^0_{\rm tot}=M$, this leads to
\begin{equation}\label{pseudoM}
M(t)=M(0)-\int^t_0\oint_S\tau^{0 j}{\rm d}^2\bar\Sigma_j{\rm d}t,
\end{equation}
where $M(0)$ can be obtained by using Eq.~(\ref{surfmass}) at $t=0$.

The total angular momentum of any isolated system as measured
gravitationally in the asymptotically flat region far from the system is
\begin{equation}
J^{\mu\nu}_{\rm tot}=2\int_V x^{[\mu}\tau^{\nu]0}{\rm d}^3 x.\label{eqn:AngTot}
\end{equation}
Thus
\begin{align}
\frac{{\rm d}J^{\mu\nu}_{\rm tot}}{{\rm d}t}
&=2\frac{\rm d}{{\rm d}t}\int_V x^{[\mu}\tau^{\nu]0}{\rm d}^3 x
=2\int_V(x^{[\mu}\tau^{\nu]0})_{,0}{\rm d}^3 x=2\int_V(\delta^{[\mu}_0
\tau^{\nu]0}+x^{[\mu}\tau^{\nu]0}{}_{,0}){\rm d}^3 x \nonumber\\
&=2\int_V[\tau^{[\nu\mu]}-(x^{[\mu}\tau^{\nu]j})_{,j}]{\rm d}^3 x
=-\oint_S(x^\mu\tau^{\nu j}-x^\nu\tau^{\mu j}){\rm d}^2\bar\Sigma_j,
\label{eqn:dAngTotdt}
\end{align}
Therefore, the total angular momentum flux across the 2-surface $S$ within
$[t_1,t_2]$ is
\begin{equation}
\Delta J^{\mu\nu}_{\rm tot}\equiv J^{\mu\nu}_{\rm tot}(t_2)-
J^{\mu\nu}_{\rm tot}(t_1)=-\int^{t_2}_{t_1}\oint_S(x^\mu\tau^{\nu j}-
x^\nu\tau^{\mu j}){\rm d}^2\bar\Sigma_j{\rm d}t,
\end{equation}
and for $J^z=\epsilon_{xy}{}^z J^{xy}$,
\begin{equation}\label{pseudoJ}
J^z(t)=J^z(0)-\int^t_0\oint_S(x\tau^{2 j}-y\tau^{1 j})
{\rm d}^2\bar\Sigma_j{\rm d}t.
\end{equation}
where $J^z(0)$ can be obtained by using Eq.~(\ref{surfang}) at $t=0$.

We will use Eqs.\,(\ref{pseudoM}) and (\ref{pseudoJ}) to calculate the
mass and the angular momentum, and compare them with the result from
Eqs.\,(\ref{surfmass}) and (\ref{surfang}).
\section{Numerical Result}\label{secIV}
\begin{figure}[thbp]
\begin{tabular}{rl}
\includegraphics[width=0.45\textwidth]{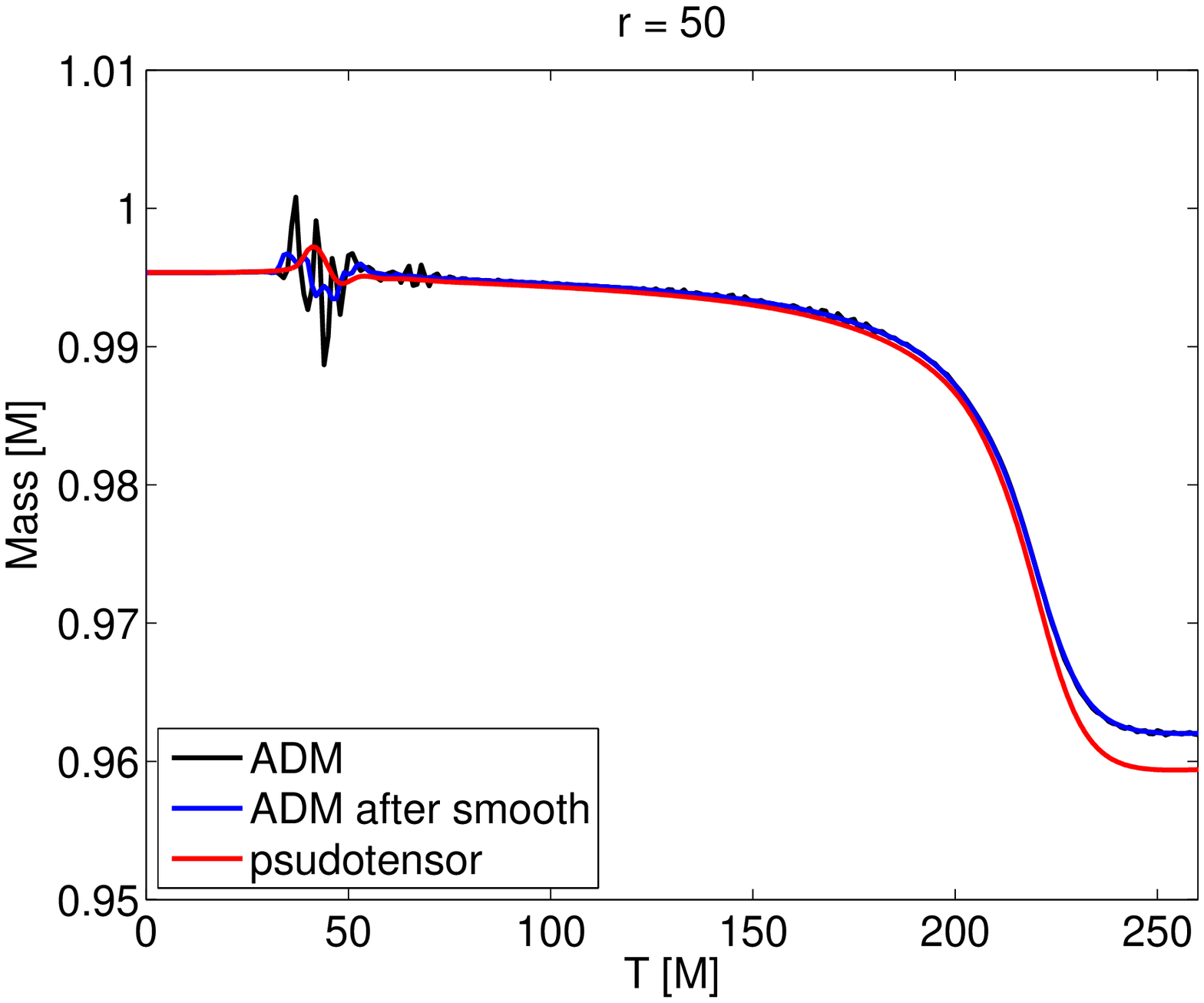}&
\includegraphics[width=0.45\textwidth]{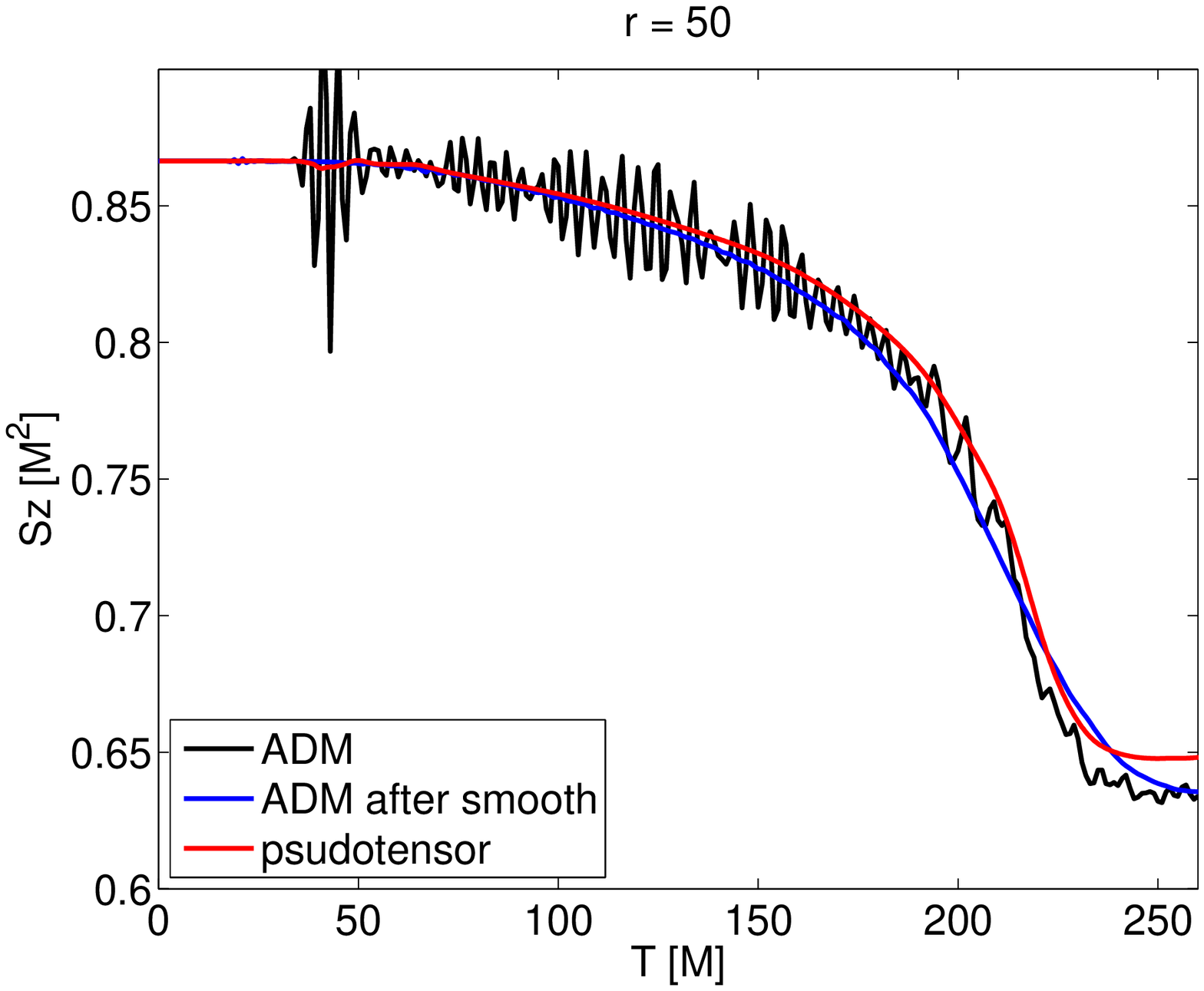}\\
\includegraphics[width=0.45\textwidth]{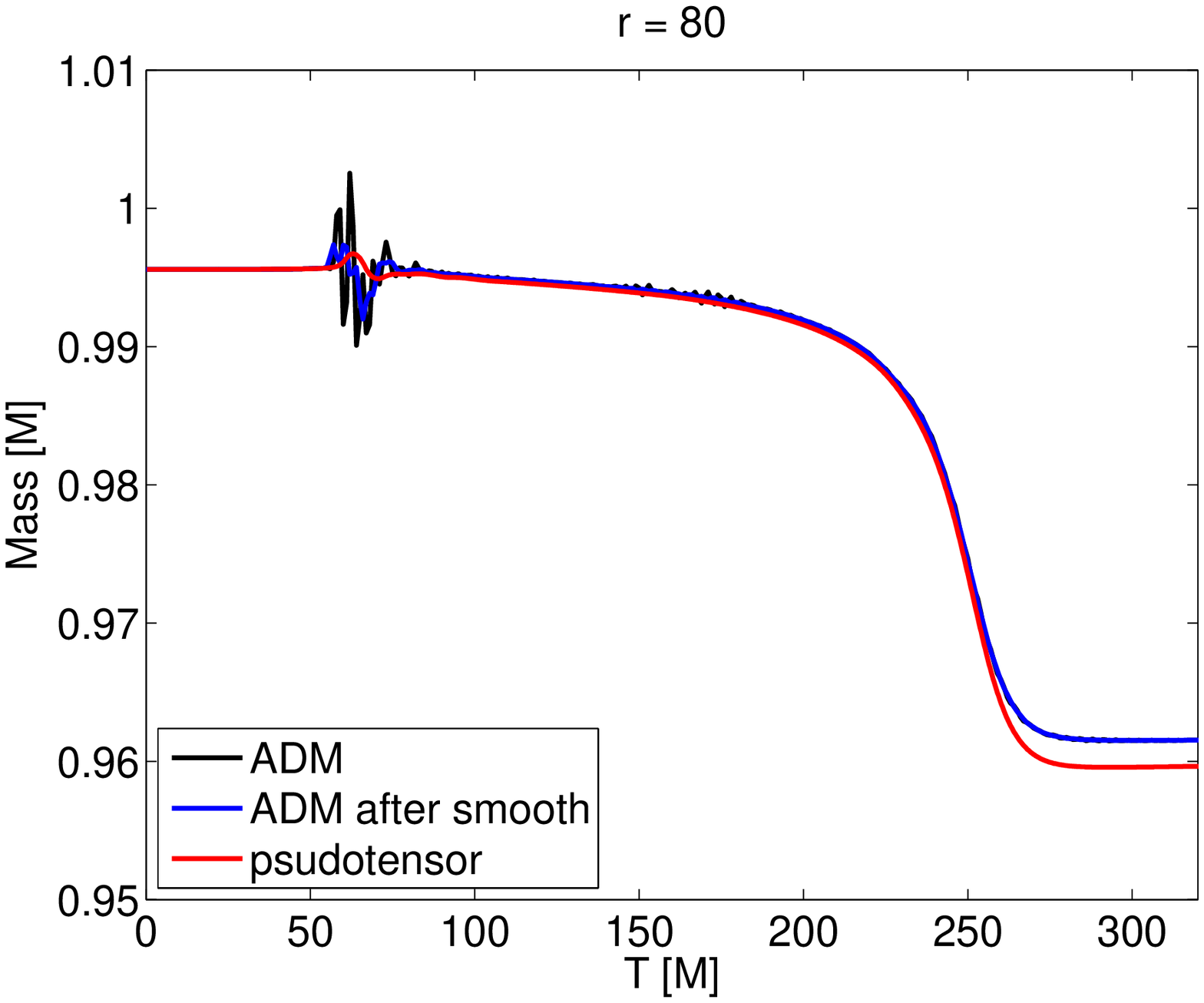}&
\includegraphics[width=0.45\textwidth]{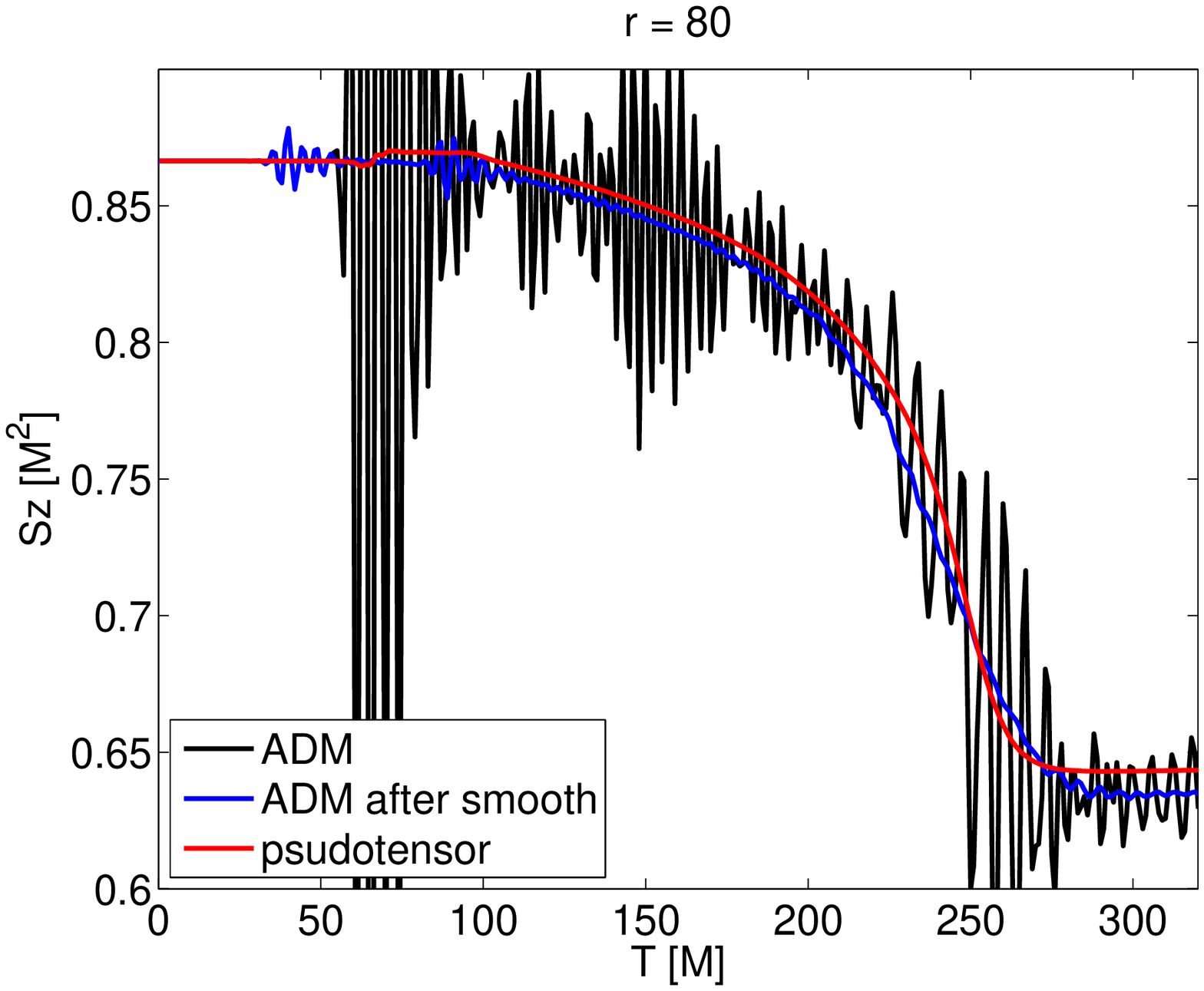}\\
\includegraphics[width=0.45\textwidth]{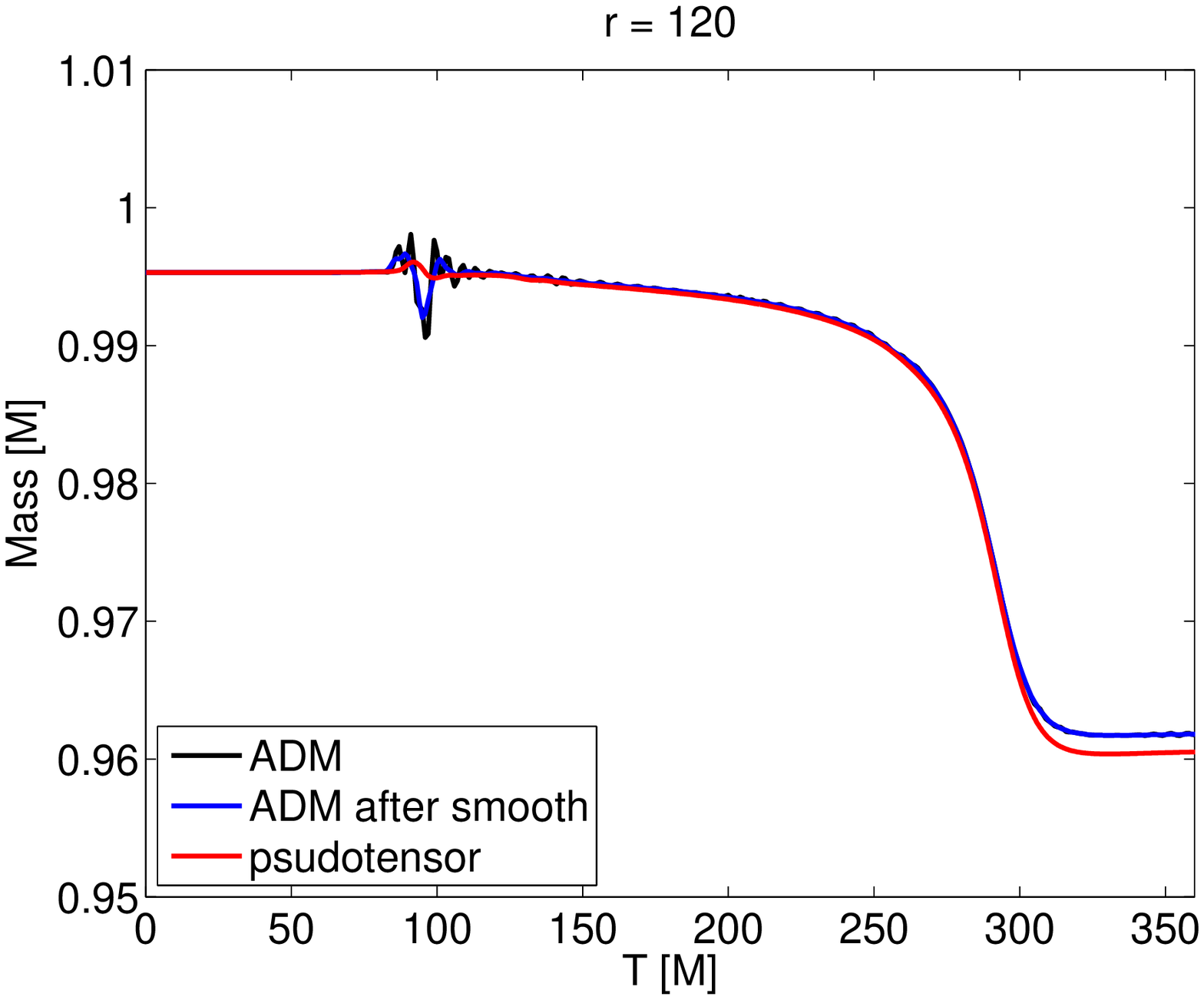}&
\includegraphics[width=0.45\textwidth]{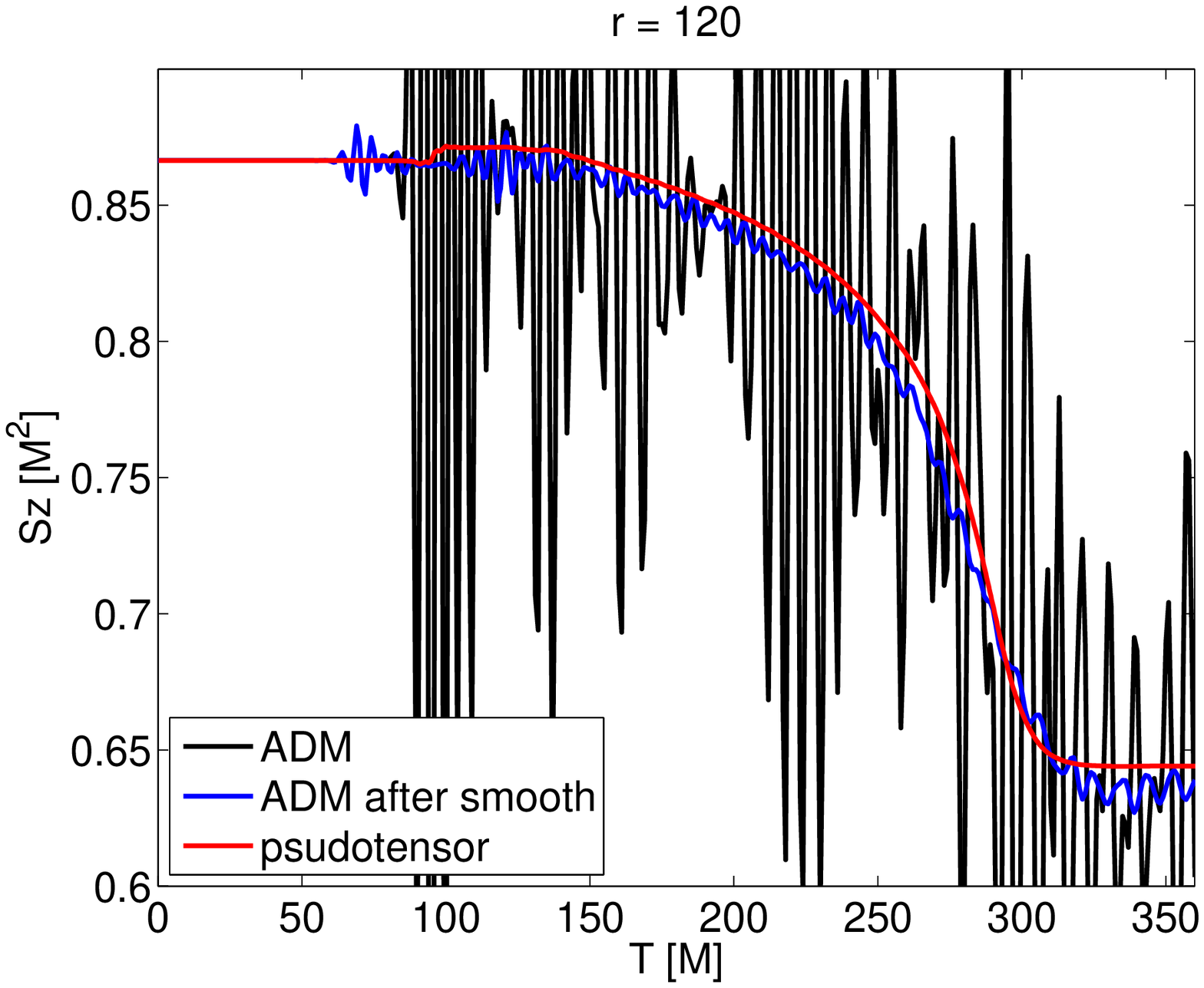}
\end{tabular}
\caption{Comparisons of the physical quantities calculated with the
traditional integration and the pseudotensor flux integrations for the
spinless BBH case.
The left column corresponds to the masses $M$ measured at the radii $r=50$,
$80$, and $120$ respectively.
The right column corresponds to the angular momenta $J_z$ of the spacetime
measured at the same radii as in the left column.}
\label{fig1}
\end{figure}
The AMSS-NCKU code with the standard moving box style mesh refinement
\cite{czyy08,cao10,yhlc12} is used in this work.
We used 10 mesh levels, and the finest 3 levels are
movable in evolving the binary black holes (BBHs).
In each fixed level, we used one box with $128\times 128\times 64$ grids
with assumed equatorial symmetry.
The outermost physical boundary is $512 M$ and this makes the finest
resolution to be $h=M/64$.
For the movable levels, two boxes with $64\times 64\times 32$ grids are used
to cover each black hole.
In time direction, the Berger-Oliger numerical scheme is adopted for the
levels higher than four.

The moving puncture gauge condition
\begin{eqnarray}
&&\partial_t\alpha=\beta^i\alpha_{,i}-2\alpha K,\\
&&\partial_t\beta^i=\frac{3}{4}B^i+\beta^j\beta^i_{,j},\\
&&\partial_tB^i=\partial_t\tilde{\Gamma}^i-\eta B^i+\beta^jB^i_{,j}-
\beta^j\tilde{\Gamma}^i_{,j}.
\end{eqnarray}
is used and has been shown to give good behavior for the black hole
simulations in \cite{czyy08}.
In this paper we use $\eta=2 M$ with $M$ being the ADM mass of the given
configuration.

In this section we apply the analysis tools described in the above section
to the inspiralling binary black hole systems.
We present two cases in this paper.
One corresponds to the spinless binary black holes initially.
The another one corresponds to two fast-spinning black holes initially.
The two individual black holes in the binary are identical in the both cases.
In the fast-spinning case, the spin parameter for each black hole is $a=0.9$.
And the spin is aligned to the orbital angular momentum.
For the detailed description of the initial data construction, the grid
setting for the numerical evolution, and the involved numerical tricks,
we refer our reader to \cite{yclp15}.
\begin{figure}[thbp]
\begin{tabular}{rl}
\includegraphics[width=0.45\textwidth]{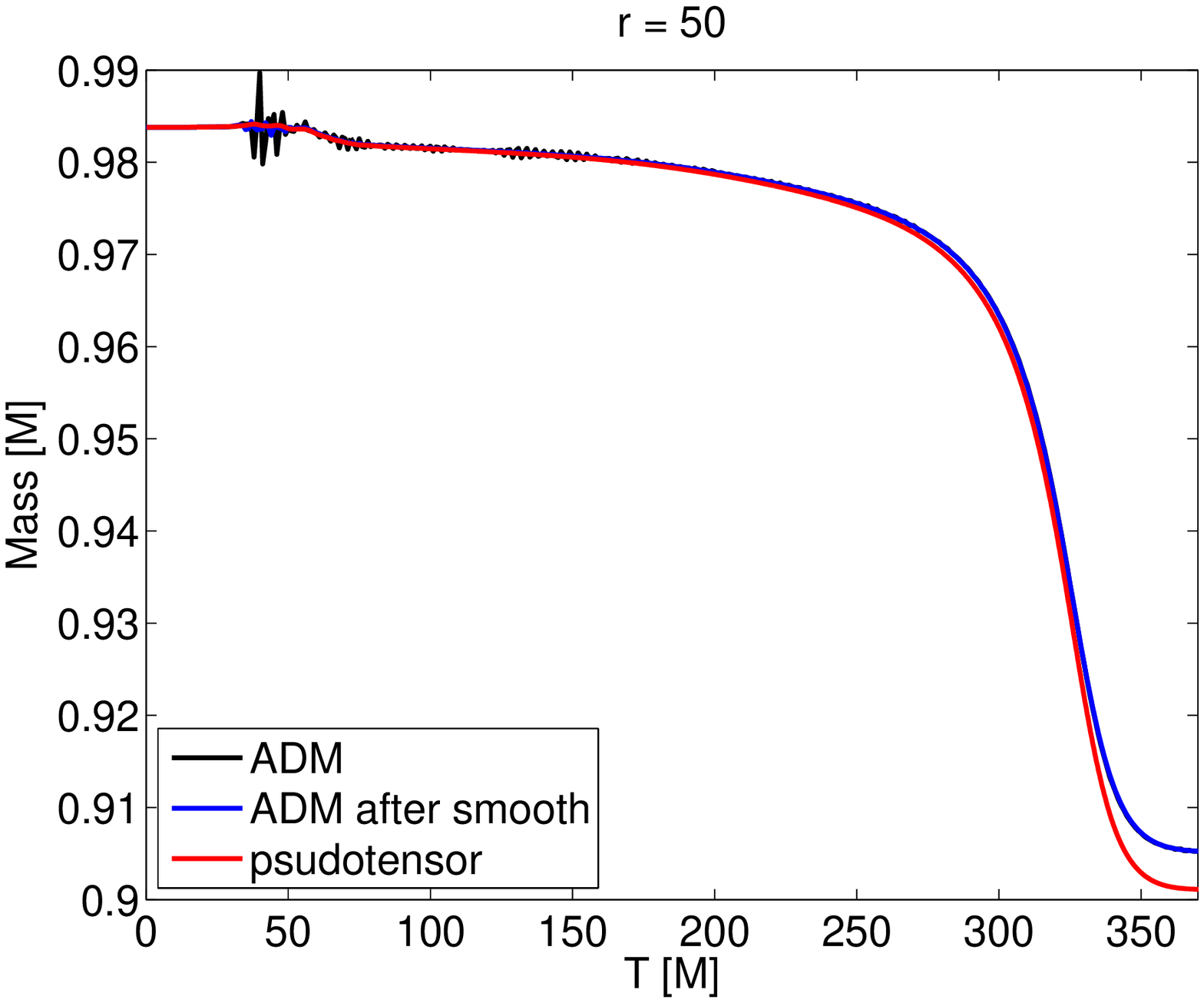}&
\includegraphics[width=0.45\textwidth]{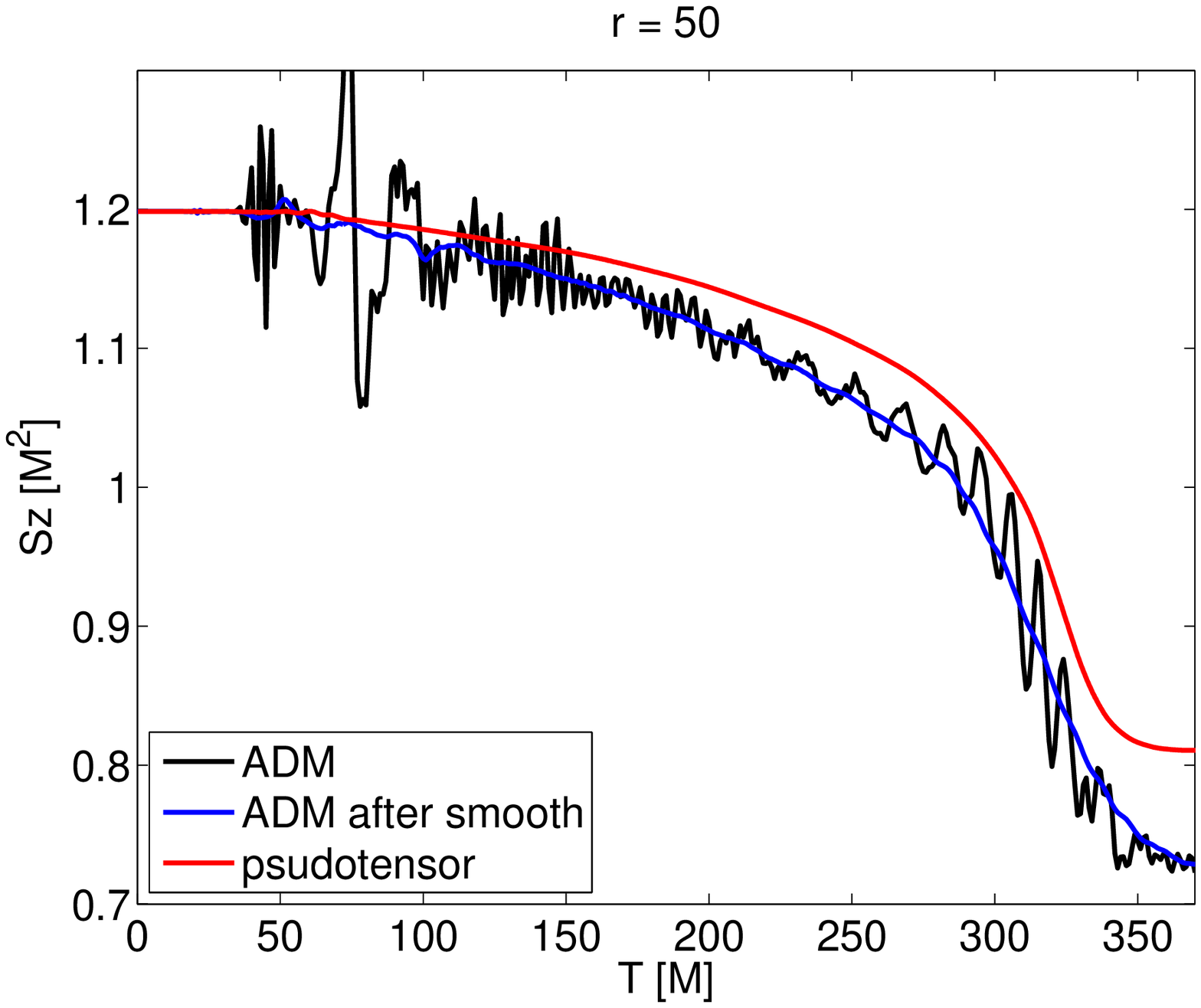}\\
\includegraphics[width=0.45\textwidth]{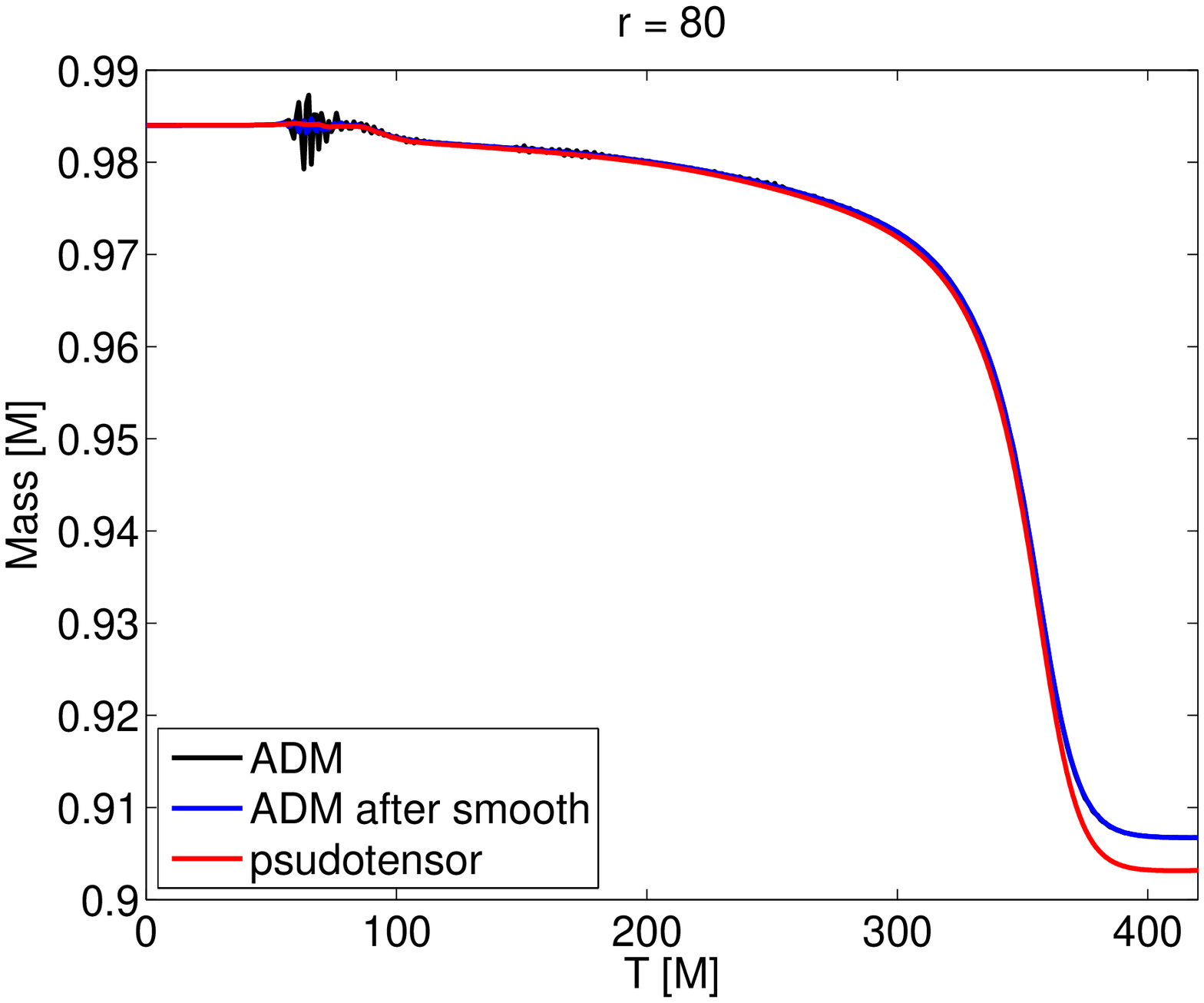}&
\includegraphics[width=0.45\textwidth]{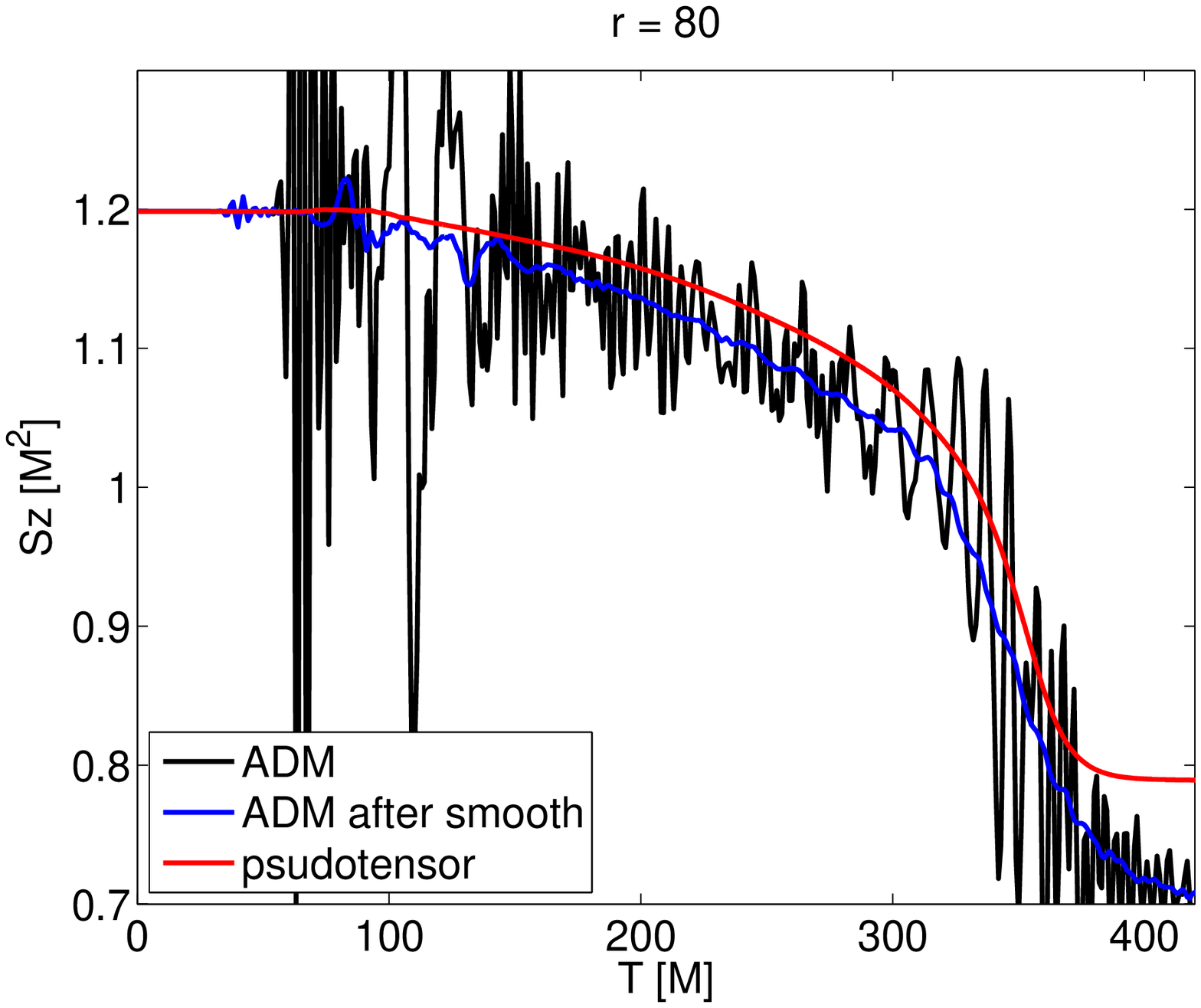}\\
\includegraphics[width=0.45\textwidth]{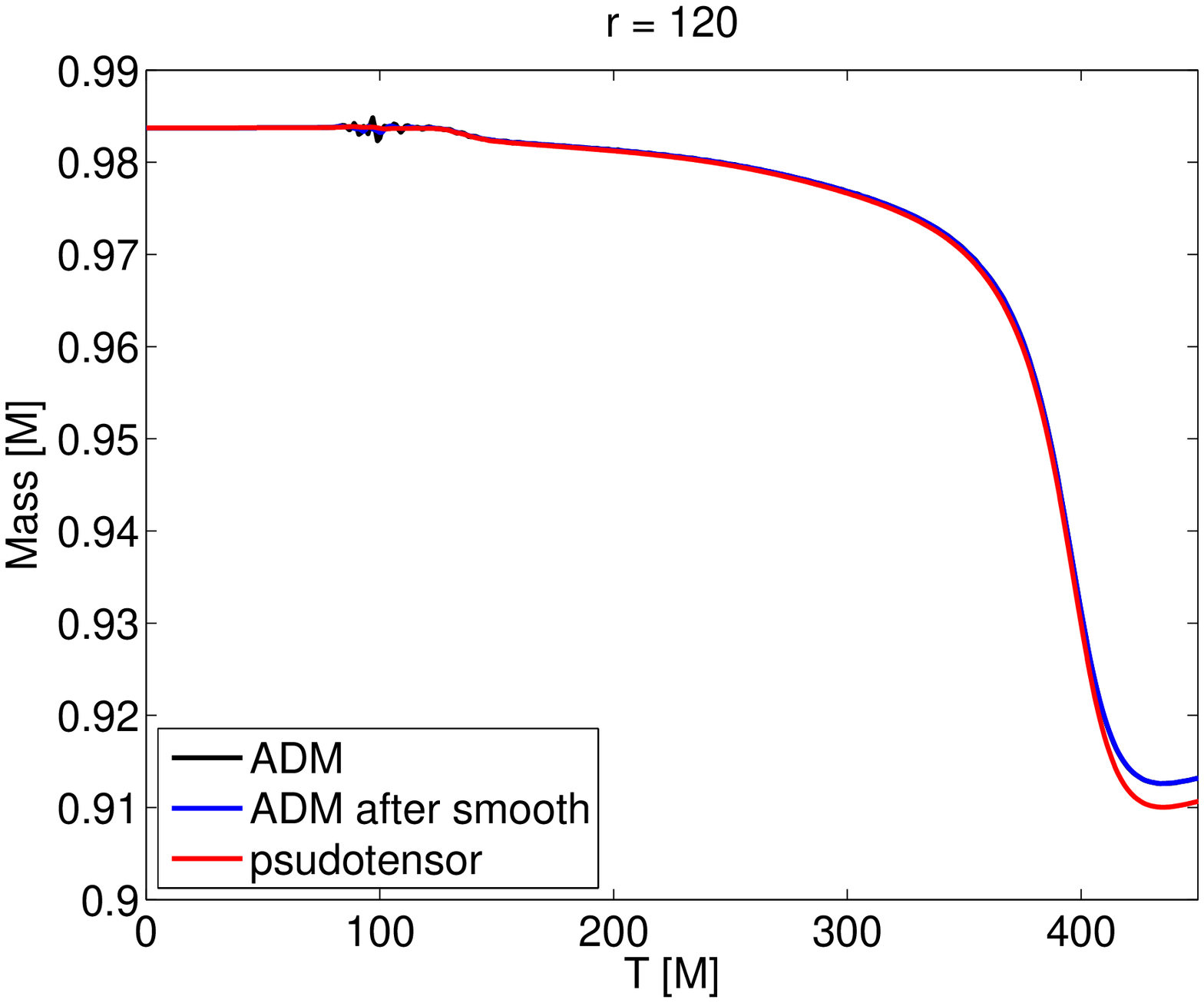}&
\includegraphics[width=0.45\textwidth]{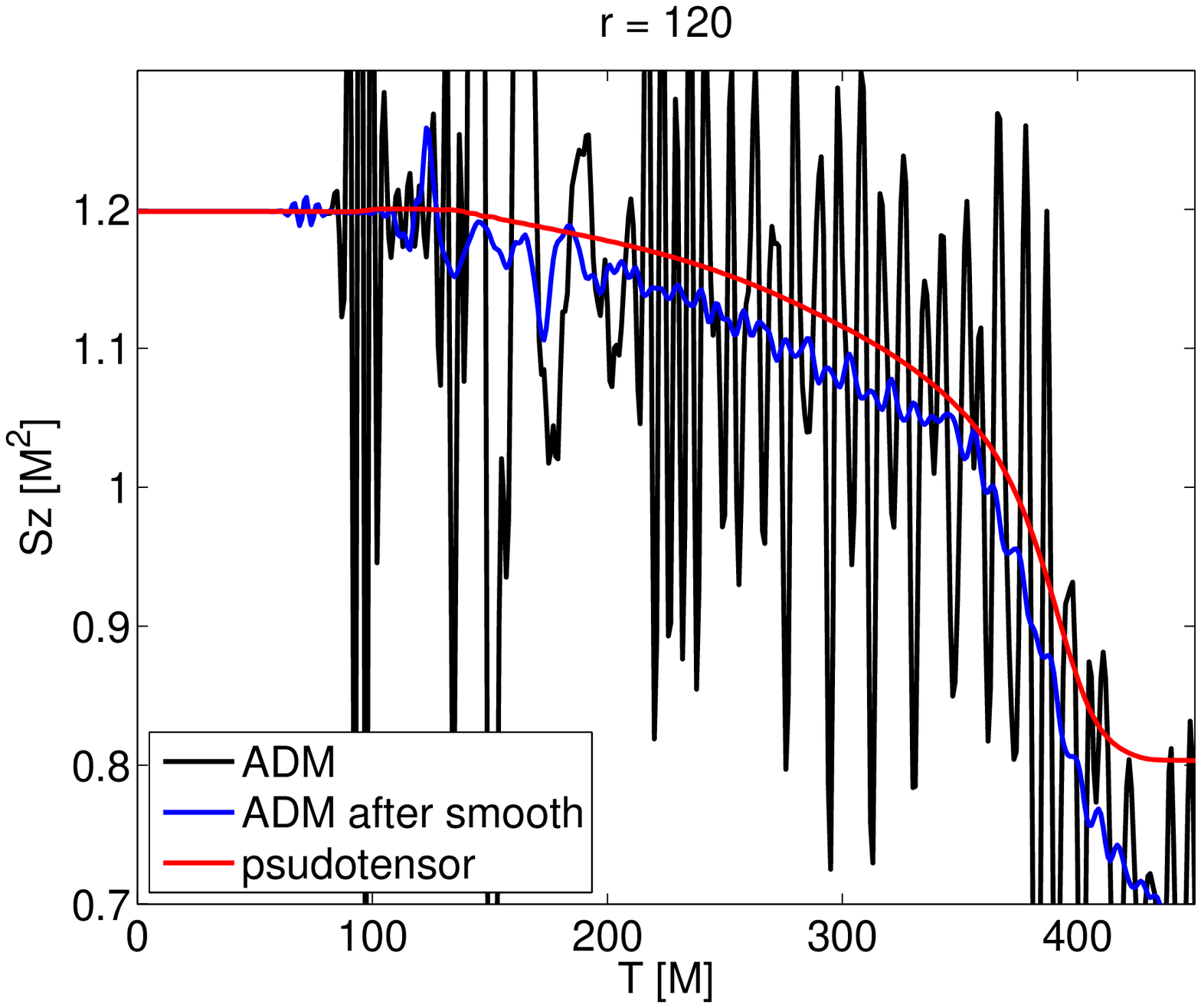}
\end{tabular}
\caption{Same as in Fig.~\ref{fig1}, except that these plots are for the BBH
case with spin $a=0.9$.}
\label{fig2}
\end{figure}
\begin{figure}[th]
\begin{tabular}{cc}
\includegraphics[width=0.45\textwidth]{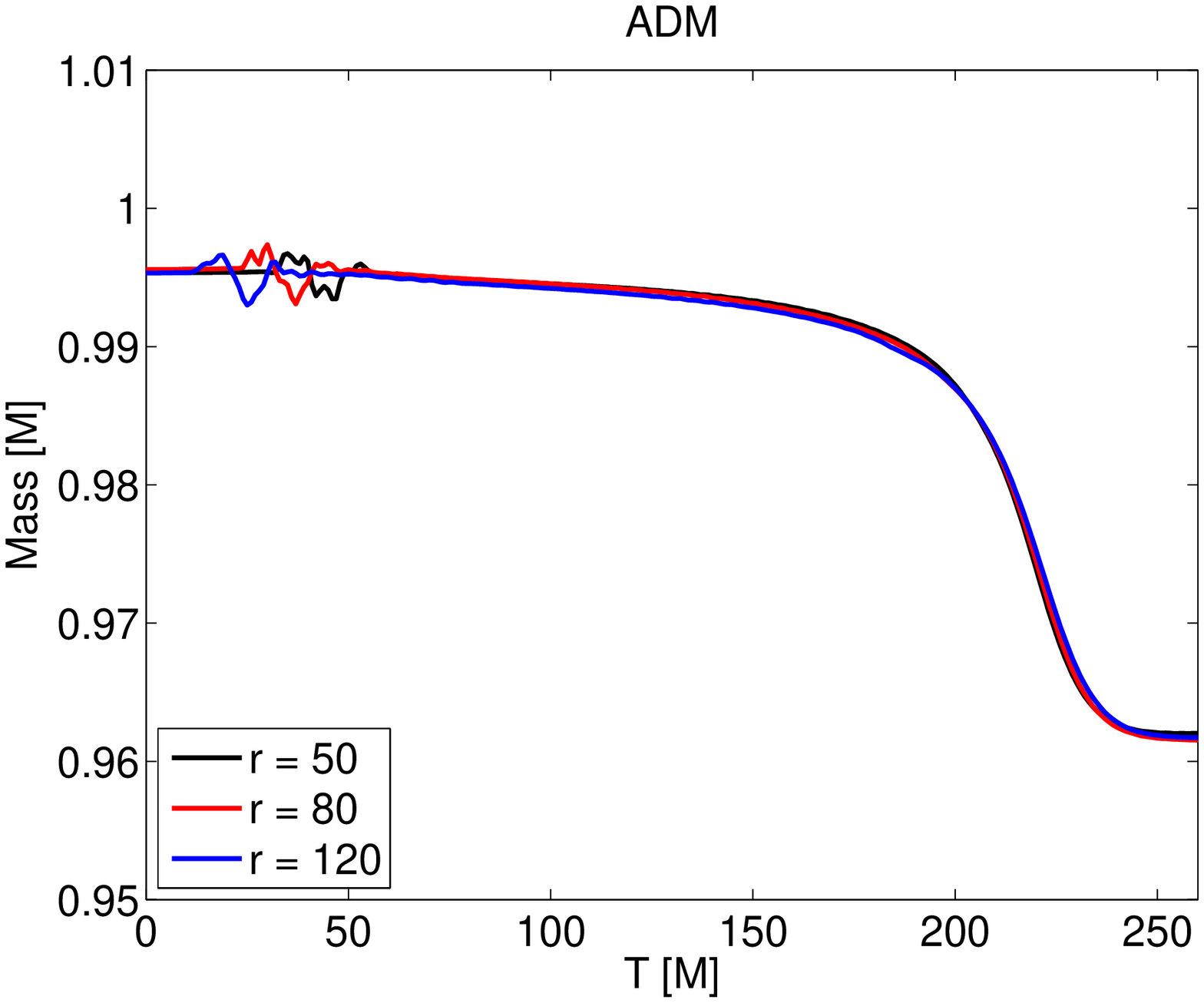}&
\includegraphics[width=0.45\textwidth]{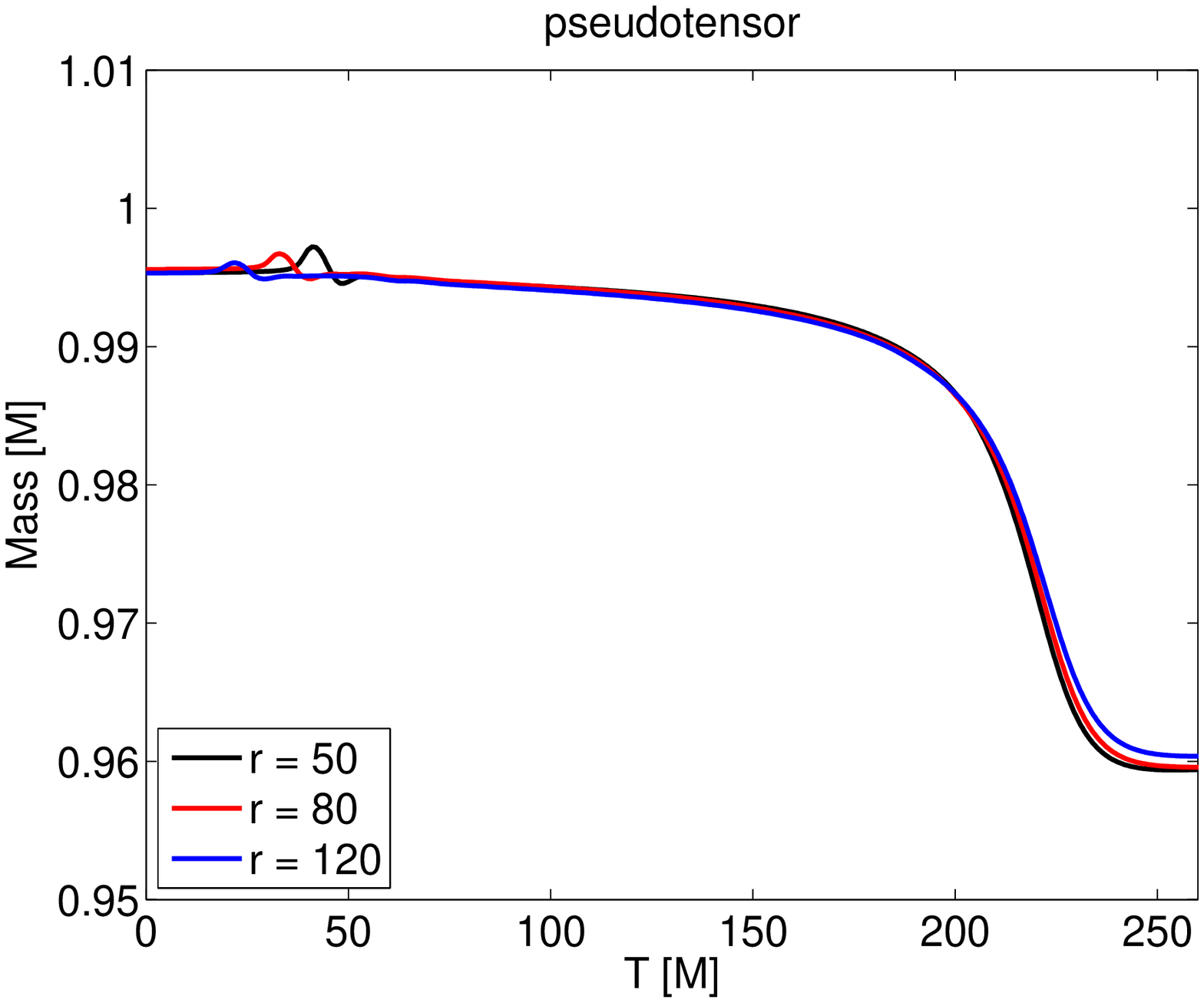}\\
\includegraphics[width=0.45\textwidth]{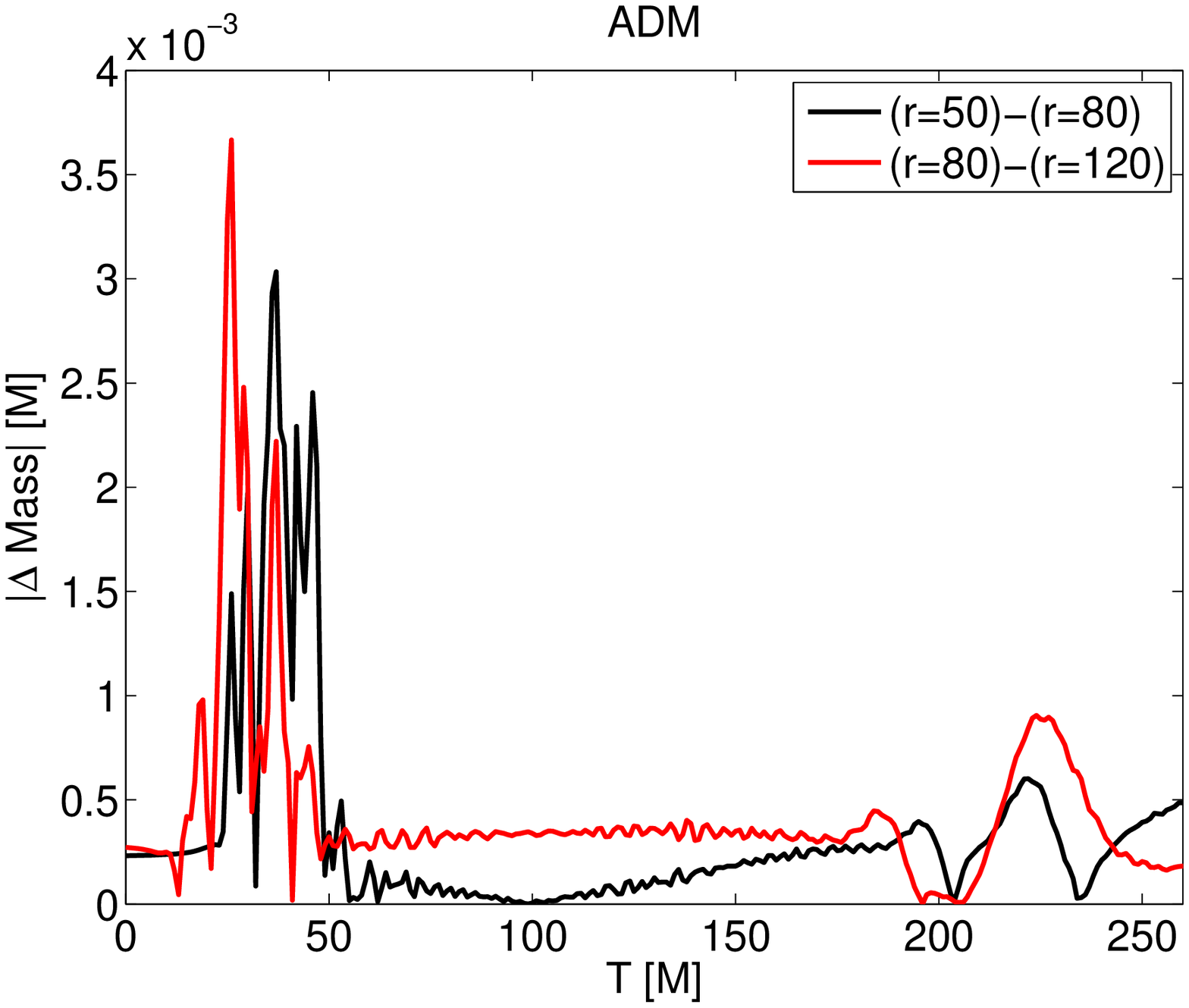}&
\includegraphics[width=0.45\textwidth]{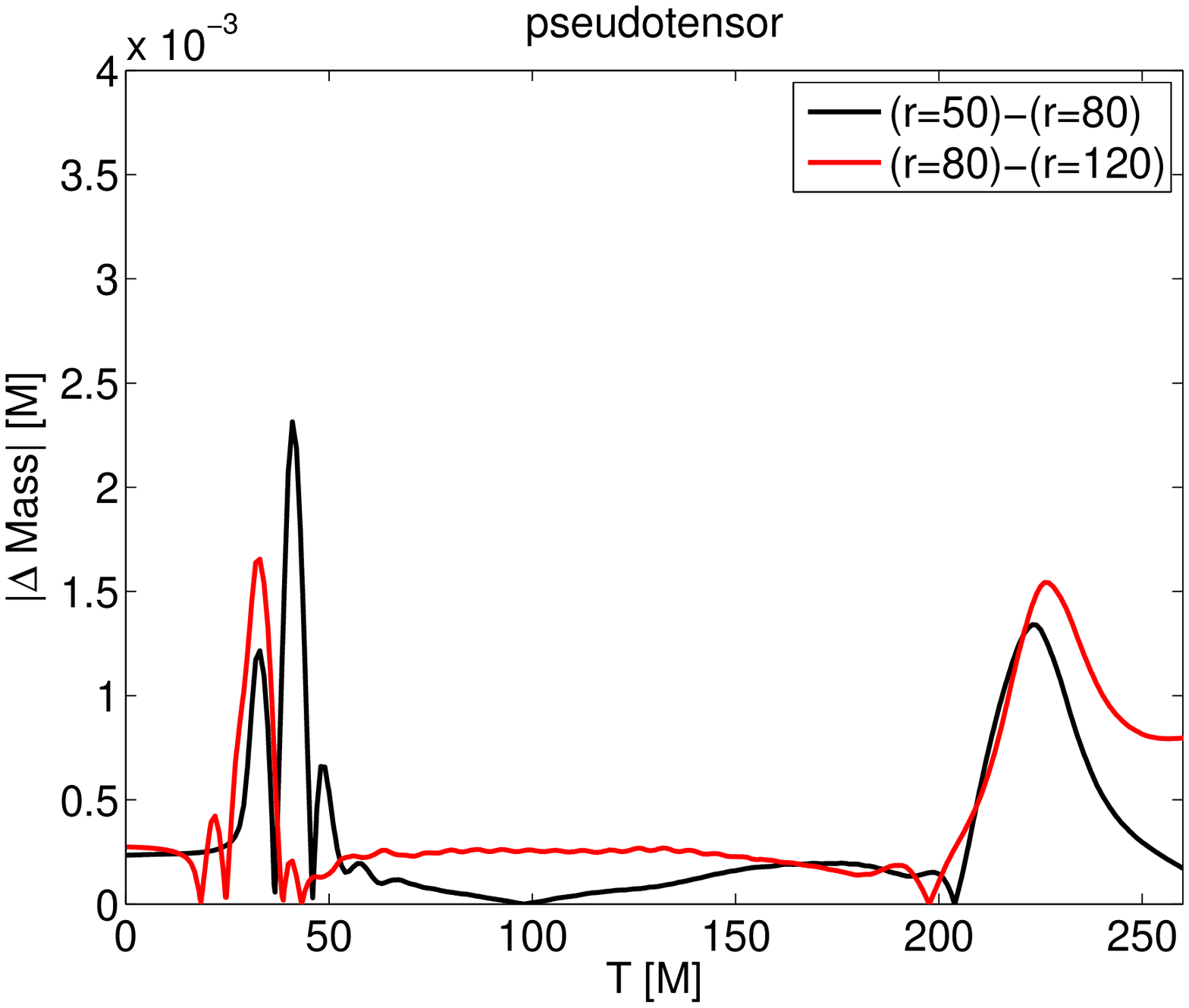}
\end{tabular}
\caption{Comparison of the effect of the finite extraction radius on
the value of mass calculated with the traditional integration and the
pseudotensor flux integration for the spinless BBH case.
The data for the ADM masses here have been smoothed as explained in
Fig.~\ref{fig1}.}
\label{fig3}
\end{figure}
\begin{figure}[th]
\begin{tabular}{cc}
\includegraphics[width=0.45\textwidth]{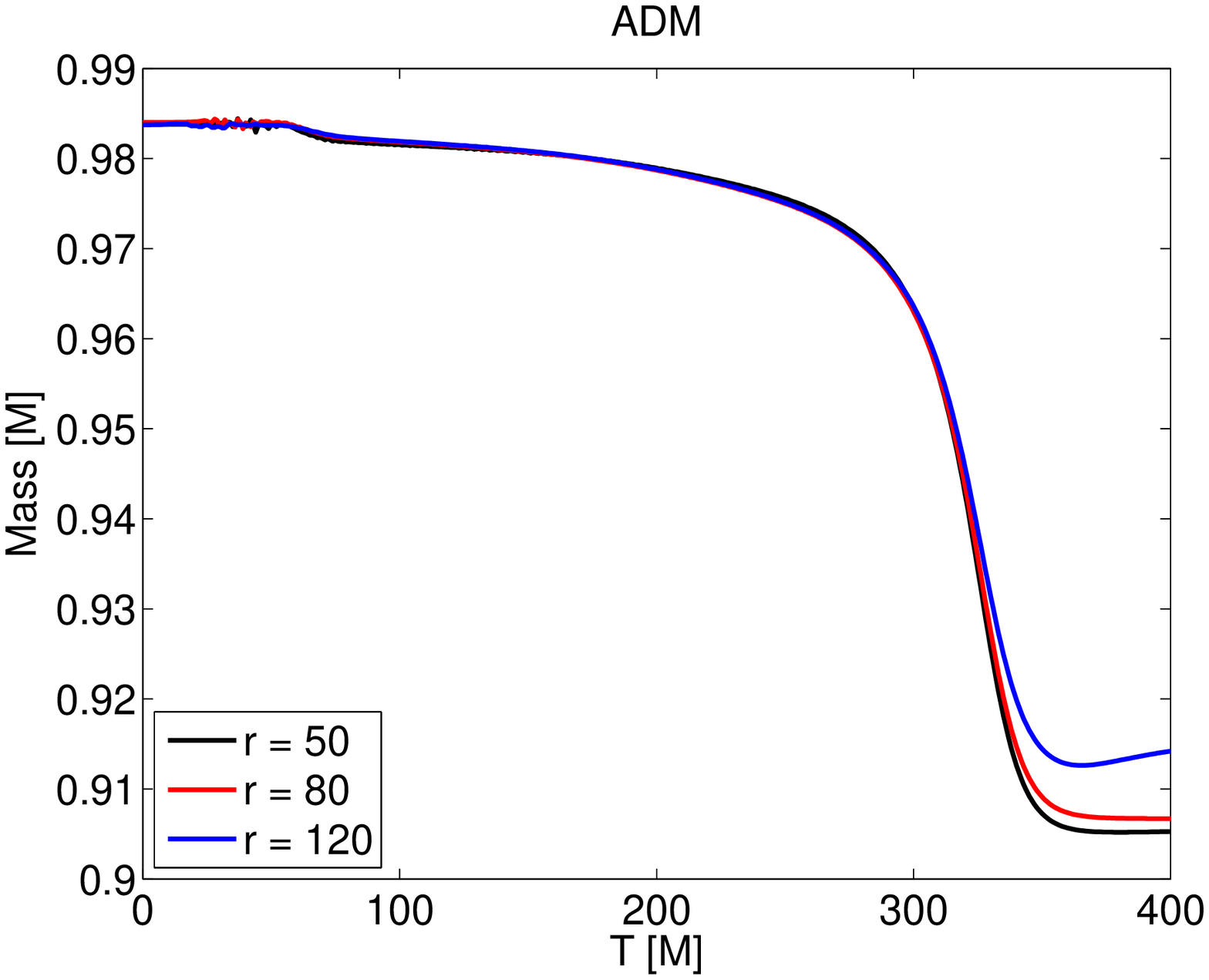}&
\includegraphics[width=0.45\textwidth]{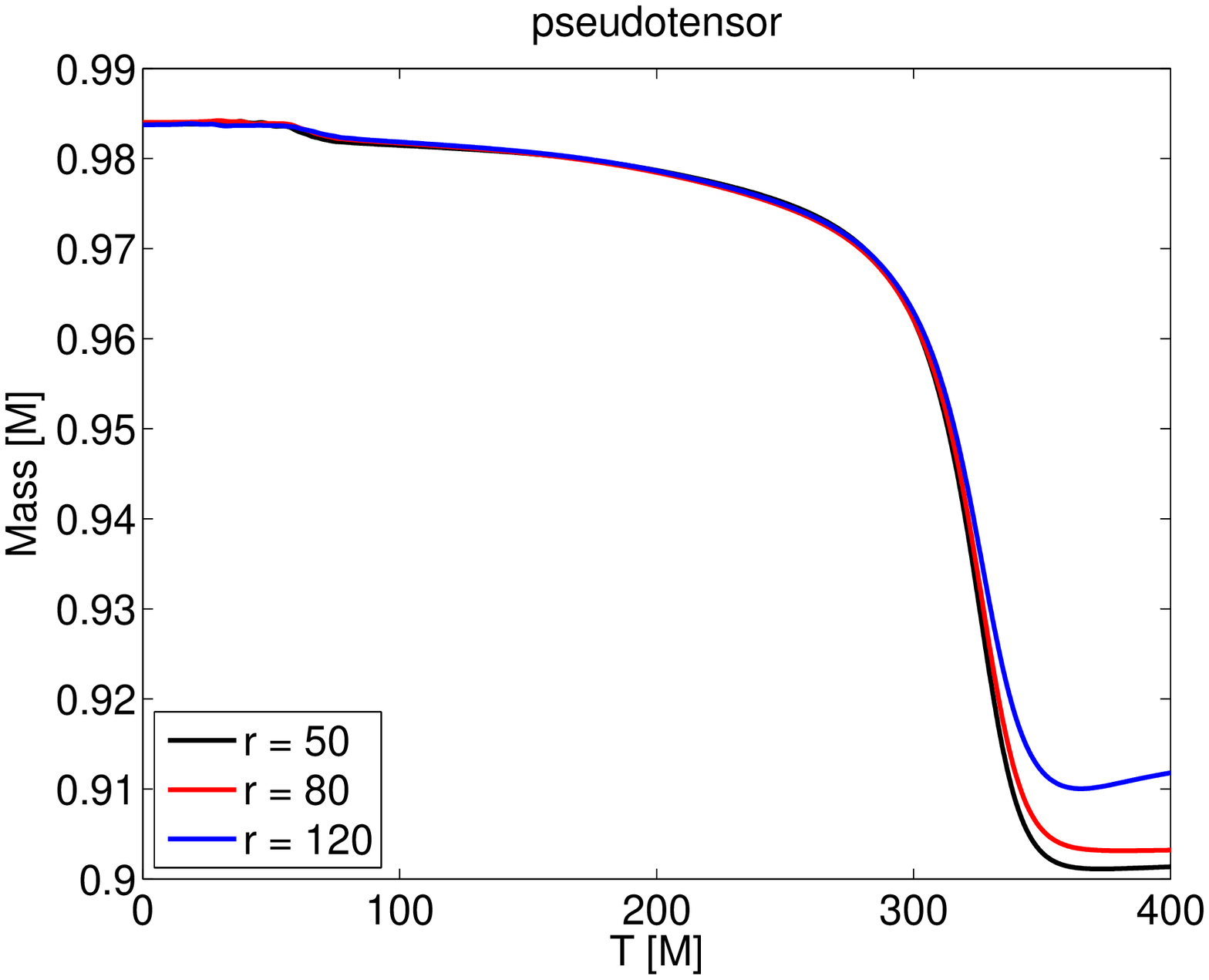}\\
\includegraphics[width=0.45\textwidth]{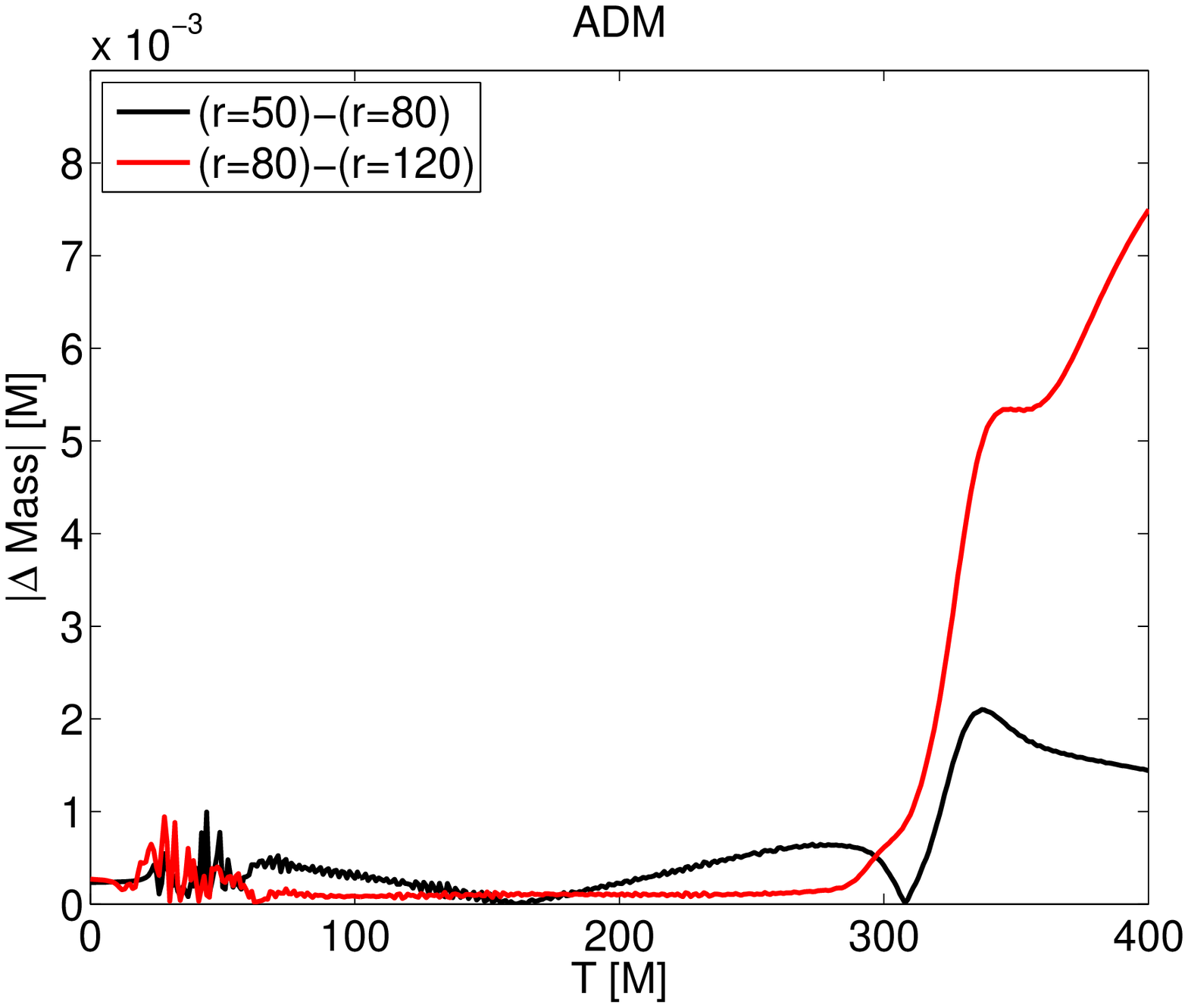}&
\includegraphics[width=0.45\textwidth]{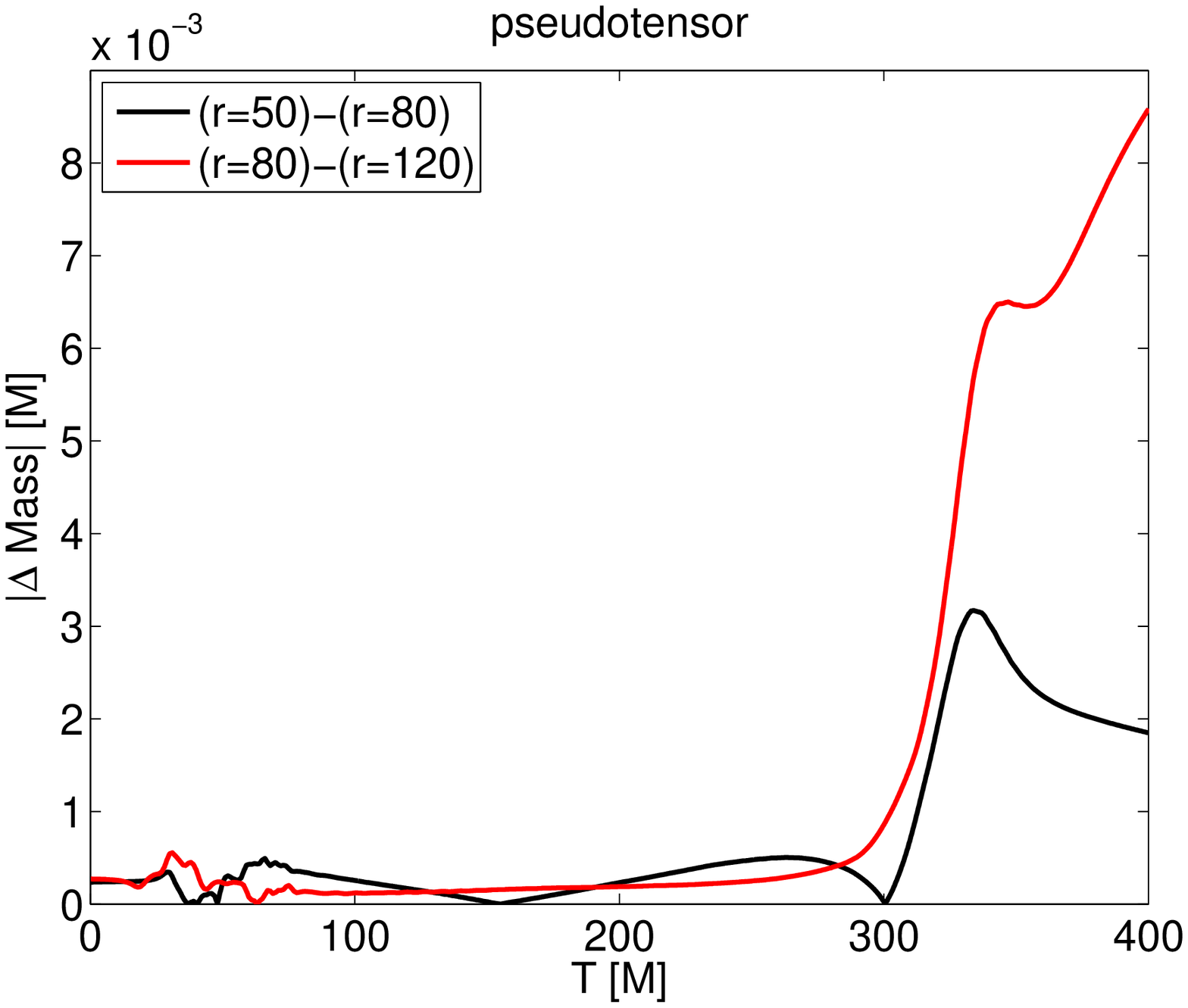}
\end{tabular}
\caption{Same as in Fig.~\ref{fig3}, except that these plots are for the
BBH case with spin $a=0.9$.}
\label{fig4}
\end{figure}

In Fig.~\ref{fig1} we compare the binary's mass and its angular momentum
calculated with the {\it traditional} integrations, i.e.,
Eqs.~(\ref{surfmass}) and (\ref{surfang}), and the pseudotensor flux
integrations, i.e, Eqs.~(\ref{pseudoM}) and (\ref{pseudoJ}).
We show the results for three different extraction radii $r=50$, $80$, and
$120$ respectively.
In the both BBH cases, the physical quantities calculated with the traditional 
integration allow larger fluctuations which come from the numerical error.
For the convenience of comparison, we smooth the traditional data by
averaging within each time range $5M$.
In the figure, we denote the data as ``ADM after smooth''.
And the data marked with ``ADM'' corresponds to the raw data.
The data after smoothing becomes much smoother.
However, by the comparison with the quantities from the pseudotensor flux
integrations, the smoothed data still fluctuates more.
In the plot of mass, such fluctuation appears after the junk radiation
reaches the extraction sphere.
We consider such fluctuation as the gauge adjustment resulted from the
junk radiation.
The numerical error could also come from the reflection of the junk radiation
via the mesh refinement boundary, and thus contribute to the fluctuation.
As we can see from Fig.~\ref{fig1}, for the traditional integration,
the angular momentum is even more sensitive to these factors.
So even after the junk radiation passes away, the angular momentum still
fluctuates mildly due to the numerical error from the mesh refinement boundary
reflection.
Interestingly, the quantities calculated with the pseudotensor flux
integrations seem immune to these factors.
Figure \ref{fig2} gives the similar result, but for fast-spinning BBH case.
For the fast-spinning BBH case, the gauge dynamics is more complicated.
So we can see the fluctuation of the traditional integration is more drastic
than the spinless case.
The result from the pseudotensor flux calculation still works smoothly
in this extreme configuration.
Except those fluctuation in the traditional integration, the results of these
two analysis tools are consistent to each other in both Fig.~\ref{fig1} and 
~\ref{fig2}.

The global quantities are formally defined at infinity.
However, we can only calculate them at some finite radius in practice.
This may cause some ambiguity.
In principle, the sequence corresponding to different extraction
radii should converge to the quantities defined at infinity.
To be a good analysis tool, we expect that the method gives a fast convergence.
In Fig.~\ref{fig3} we compare the convergence behavior of the mass integral
with respect to different extraction radii with the traditional integration
and the pseudotensor flux integration for the spinless BBH case.
During the junk radiation period, the convergence of the pseudotensor flux
method is roughly two times better than the traditional integration.
But during the merger part, the convergence of the traditional integration
method is two times better than the pseudotensor flux method.
Considering that the junk radiation is unphysical, we conclude that
the traditional integration method is a better analysis tool in this aspect.
In Fig.~\ref{fig4}, we did the same investigation for fast-spinning BBH case.
For the junk radiation part the same result can be seen as the spinless case.
For the merger part, the fast-spinning BBH configuration introduces some
challenge to the numerical evolution as explained in \cite{yclp15}.
So as ones expect, the convergence behaviors for both analysis methods are
equally bad, although they are consistent with each other.
As to the angular momentum, the result from the traditional integration
fluctuate so much that it does not make sense to compare it with the one
from the pseudotensor flux integration.
\section{Summary}\label{secV}
In this work we apply the Landau-Lifshitz pseudotensor flux formalism as
an alternative method in calculating the total mass and the total angular
momentum during the evolutions of a binary black hole system.
We also compare its performance with the traditional integrations for the 
global quantities.
Due to the gauge choice employing a flat spacetime background in the
Landau-Lifshitz pseudotensor formalism,
it is not expected that the result from this method will be accurate
enough for the radius for integration is not far from the singularity.
However, we find that the overall result with the method is consistent
with the one with the traditional integration.

The advantage of the pseudotensor flux formalism is the smoothness of the 
global quantities, especially of the total angular momentum. 
It has been plagued for a long time with the fluctuation and inaccuracy of
the numerical value of the total angular momentum calculated with the
traditional integration, especially when the grid resolution is usually low
for the radius of the surface integration is large.
It shows in this work that this problem can be solved with the pseudotensor
flux method.
The reason mainly comes from the integrations along the time domain in 
Eqs.~(\ref{pseudoM}) and (\ref{pseudoJ}).
Therefore, although the convergence behavior of the global quantities with 
the pseudotensor flux method is only comparable with the ones with 
the traditional method, the smoothness of its numerical value allows using
a larger radius for surface integration to obtain more accurate result.

As showed in \cite{mpea08} and \cite{yclp15}, the total angular momentum
calculated with the traditional method usually decays after the merger
in the fast-spinning BBH cases.
In our BBH simulations, it seems that the total angular momentum with the 
pseudotensor flux method conserves much better than the one with the 
traditional method. However, it might need a further detailed investigation
to confirm this point.

This work shows that the pseudotensors (and the quasi-local quantities) could
be very useful analysis tools in numerical relativity.
Therefore, we plan to study the usefulness of different
pseudotensors/quasi-local quantities, and also the advantage of different 
spacetime background, e.g, the Schwarzschild spacetime or the Kerr spacetime,
in numerical relativity in the future.
\section*{Acknowledgments}
CCT is grateful to Prof.~Shih-Yuin Lin and Prof.~I-Ching Yang for their
useful discussions.
This work was supported in part by the National Science Council under Grants
No.~NSC102-2112-M-006-014-MY2, by the Ministry of Science and Technology
under Grant No.~MOST104-2112-M-006-020,
and by the Headquarters of University Advancement at the National Cheng Kung
University, which is sponsored by the Ministry of Education, Taiwan, ROC.
Z.~Cao was supported by the NSFC (No.~11375260).
We are grateful to the National Center for High-performance Computing
for the use of their computer time and facilities. We are also
grateful to the Academia Sinica Computing Center for providing
computing resource.
\appendix
\section{Spacetime 4-connection in 3+1 Expression}\label{app1}
The 4-metric $g_{\mu\nu}$ can be constructed out of the 3-metric $\gamma_{ij}$
and the lapse $\alpha$ and shift functions $\beta^i$ as
\begin{align}
\left\Vert
\begin{array}{ccc}
g_{00}&&g_{0 k}\\&&\\g_{j 0}&&g_{jk}
\end{array}
\right\Vert &=
\left\Vert
\begin{array}{ccc}
\beta^\ell\beta_\ell-\alpha^2&&\beta_k\\&&\\\beta_j&&\gamma_{jk}
\end{array}
\right\Vert,\\
\left\Vert
\begin{array}{ccc}
g^{00}&&g^{0 k}\\&&\\g^{j 0}&&g^{jk}
\end{array}
\right\Vert &=
\left\Vert
\begin{array}{ccc}
-\displaystyle\frac{1}{\alpha^2}&&\displaystyle\frac{\beta^k}{\alpha^2}\\&&\\
\displaystyle\frac{\beta^j}{\alpha^2}&&\gamma^{jk}-\displaystyle
\frac{\beta^j\beta^k}{\alpha^2}
\end{array}
\right\Vert,
\end{align}
where $\beta_i=\gamma_{ij}\beta^j$.

From Appendix B of \cite{alcubierre08} we can obtain the following expressions
for the 4-connection in terms of 3+1 quantities
\begin{align}
{\mathit\Gamma}^0{}_{ij}&=-\frac{1}{\alpha}K_{ij},\\
{\mathit\Gamma}^0{}_{0i}&=\frac{1}{\alpha}(\nabla_i\alpha-K_{im}\beta^m)=
\nabla_i\ln\alpha+{\mathit\Gamma}^0{}_{im}\beta^m,\\
{\mathit\Gamma}^0{}_{00}&=\frac{1}{\alpha}(\partial_t\alpha+\beta^m
\nabla_m\alpha-K_{mn}\beta^m\beta^n)=\partial_t\ln\alpha+\beta^m
{\mathit\Gamma}^0{}_{0m},\\
{\mathit\Gamma}^i{}_{jk}&=\Gamma^i{}_{jk}+\frac{\beta^i}{\alpha}K_{jk}
=\Gamma^i{}_{jk}-\beta^i{\mathit\Gamma}^0{}_{jk},\\
{\mathit\Gamma}^i{}_{0j}&=\nabla_j\beta^i-\alpha K^i{}_j
+\frac{\beta^i}{\alpha}(K_{jm}\beta^m-\nabla_j\alpha)
=\nabla_j\beta^i-\alpha K^i{}_j-\beta^i{\mathit\Gamma}^0{}_{0j},\\
{\mathit\Gamma}^i{}_{00}&=\partial_t\beta^i+\beta^m\nabla_m\beta^i
+\alpha(\nabla^i\alpha-2K^i{}_m\beta^m)\nonumber\\
&+\frac{\beta^i}{\alpha}(K_{mn}\beta^m\beta^n-\partial_t\alpha-\beta^m
\nabla_m\alpha)\nonumber\\
&=\partial_t\beta^i+\beta^m{\mathit\Gamma}^i{}_{0m}+(\alpha^2\gamma^{im}+
\beta^i\beta^m){\mathit\Gamma}^0{}_{0m}-\beta^i{\mathit\Gamma}^0{}_{00},
\end{align}
where $\nabla_i$ is the covariant derivative associated with the 3-metric
$\gamma_{ij}$, and the corresponding 3-connection $\Gamma^i{}_{jk}$.
For ${\mathit\Gamma}^\mu\equiv g^{\lambda\sigma}{\mathit\Gamma}^\mu
{}_{\lambda\sigma}$, $L_\mu\equiv{\mathit\Gamma}^\sigma{}_{\mu\sigma}$,
\begin{align}
L_0&=\partial_t\ln\alpha+\nabla_m\beta^m-\alpha K,\\
L_i&=\partial_i\ln\alpha+\Gamma^m{}_{mi},\\
{\mathit\Gamma}^0&=\frac{1}{\alpha^3}(\beta^m\partial_m\alpha-\partial_t\alpha
-\alpha^2 K),\\
{\mathit\Gamma}^i&=\Gamma^i-\frac{1}{\alpha}(\partial^i\alpha-\beta^i K)
-\frac{1}{\alpha^2}(\partial_t\beta^i-\beta^m
\partial_m\beta^i)+\frac{\beta^i}{\alpha^3}(\partial_t\alpha-\beta^m\partial_m
\alpha),
\end{align}
where $\Gamma^i\equiv\gamma^{jk}\Gamma^i{}_{jk}$.


\begin{thebibliography}{set}
\bibitem{pref05}
F. Pretorius, Phys. Rev. Lett. {\bf 95}, 121101 (2005).

\bibitem{lali62}
L.D. Landau and E.M. Lifshitz, {\it Classical Theory of Fields},
Addison Wesley, Redding Mass., (1962).

\bibitem{szal04}
L.B. Szabados, Living Rev. Relativity {\bf 7}, 4 (2004).

\bibitem{ccnt05}
C.M. Chen, J.M. Nester, and R.S. Tung, Phys. Rev. D{\bf 72}, 104020 (2005).

\bibitem{aakb04}
A. Ashtekar and B. Krishnan, Living Rev. Relativity {\bf 7}, 10 (2004).

\bibitem{kdea09}
D. Keppel, {\it et al}, Phys.~Rev.~D \textbf{80}, 124015 (2009);
G. Lovelace, {\it et al}, ibid. \textbf{82}, 064031 (2010).

\bibitem{mvfm15}
V. Mewes, J.A. Font, and P.J. Montero, arXiv:1505.07225.

\bibitem{BSSN95_99}
M.~Shibata and T.~Nakamura, Phys.~Rev.~D \textbf{52}, 5428 (1995);
T.W.~Baumgarte and S.L.~Shapiro, ibid. {\bf 59}, 024007 (1999).

\bibitem{MTW}
C.W. Misner, K. Throne, J.A. Wheeler, {\it Gravitation}, Freeman, San
Francisco (1973).

\bibitem{mnyj74}
N. \'O Murchadha and J. W. York, Jr., Phys. Rev. D {\bf 10}, 2345, (1974).

\bibitem{york79} J. W. York, Jr., in {\it Sources of Gravitational
Radiation}, edited by L.L. Smarr (Cambridge Univ. Press, Cambridge,
1979).

\bibitem{by80} J. M. Bowen and J. W. York, Jr., Phys. Rev. D {\bf 21}
2047 (1980).

\bibitem{yhbs02}
H.J. Yo, T.W. Baumgarte, S.L. Shapiro, Phys.~Rev.~D \textbf{66}, 084026 (2002).

\bibitem{czyy08}
Z. Cao, H.J. Yo, and J.P. Yu, Phys. Rev. D {\bf 78}, 124011 (2008).

\bibitem{cao10}
P. Galaviz, B. Br\"ugmann, and Z. Cao, Phys. Rev. D {\bf 82} 024005 (2010).

\bibitem{yhlc12}
H.J. Yo, C.Y. Lin, and Z. Cao, Phys. Rev. D {\bf 86} 064027 (2012).

\bibitem{yclp15}
H.J. Yo, Z. Cao, C.Y. Lin, and H.P. Pan, Phys. Rev. D {\bf 92} 024034 (2015).

\bibitem{mpea08}
P. Marronetti, W. Tichy, B. Bruegmann, J. Gonzalez, and U. Sperhake,
Phys. Rev. D {\bf 77}, 064010 (2008).

\bibitem{alcubierre08}
M. Alcubierre, {\it Inttroduction to $3+1$ Numerical Relativity},
Oxford  University Press (2008).

\end{thebibliography}
\end{document}